\documentclass[letterpaper]{article} % DO NOT CHANGE THIS
\usepackage{aaai2026}  % DO NOT CHANGE THIS
\usepackage{times}  % DO NOT CHANGE THIS
\usepackage{helvet}  % DO NOT CHANGE THIS
\usepackage{courier}  % DO NOT CHANGE THIS
\usepackage[hyphens]{url}  % DO NOT CHANGE THIS
\usepackage{graphicx} % DO NOT CHANGE THIS
\usepackage{natbib}  % DO NOT CHANGE THIS AND DO NOT ADD ANY OPTIONS TO IT
\usepackage{caption} % DO NOT CHANGE THIS AND DO NOT ADD ANY OPTIONS TO IT

\usepackage{array}

\usepackage{textcomp}
\usepackage{amssymb}
\usepackage{verbatim}

\usepackage{colortbl}
\usepackage{xcolor} 
\usepackage{booktabs}
\usepackage{multirow}
\usepackage{comment}
\usepackage{subcaption}

\usepackage{microtype}

\usepackage{mathtools}
\usepackage{amsthm}
\usepackage{threeparttable}

\usepackage[capitalize,noabbrev]{cleveref}
\usepackage[textsize=tiny]{todonotes}
\definecolor{lightblue}{RGB}{230, 240, 255}
\definecolor{lightgray}{gray}{0.9}

\theoremstyle{plain}

\theoremstyle{definition}

\theoremstyle{remark}

\usepackage{algorithm}
\usepackage{algorithmic}

\usepackage{newfloat}
\usepackage{listings}
\DeclareCaptionStyle{ruled}{labelfont=normalfont,labelsep=colon,strut=off} % DO NOT CHANGE THIS
\floatstyle{ruled}
\newfloat{listing}{tb}{lst}{}
\floatname{listing}{Listing}
\title{Reinforced Rate Control for Neural Video Compression via Inter-Frame Rate-Distortion Awareness}
\author{
    Wuyang Cong\textsuperscript{\rm 1},
    Junqi Shi\textsuperscript{\rm 1},
    Lizhong Wang\textsuperscript{\rm 2},
    Weijing Shi\textsuperscript{\rm 2},
    Ming Lu\textsuperscript{\rm 1}\thanks{Corresponding authors.},
    Hao Chen\textsuperscript{\rm 1},
    and Zhan Ma\textsuperscript{\rm 1$\ast$}
}
\affiliations{
    \textsuperscript{\rm 1}School of Electronic Science and Engineering, Nanjing University \\ \textsuperscript{\rm 2}Samsung Research China-Beijing\\
    \{congwuyang,junqishi\}@smail.nju.edu.cn, \{lz.wang,weijing\_.shi\}@samsung.com, \{minglu,chenhao1210,mazhan\}@nju.edu.cn
}

\usepackage{bibentry}
\begin{document}

\maketitle

\begin{abstract}
Neural video compression (NVC) has demonstrated superior compression efficiency, yet effective rate control remains a significant challenge due to complex temporal dependencies. Existing rate control schemes typically leverage frame content to capture distortion interactions, overlooking inter-frame rate dependencies arising from shifts in per-frame coding parameters. This often leads to suboptimal bitrate allocation and cascading parameter decisions. To address this, we propose a reinforcement-learning (RL)-based rate control framework that formulates the task as a frame-by-frame sequential decision process. At each frame, an RL agent observes a spatiotemporal state and selects coding parameters to optimize a long-term reward that reflects rate-distortion (R-D) performance and bitrate adherence. Unlike prior methods, our approach jointly determines bitrate allocation and coding parameters in a single step, independent of group of pictures (GOP) structure. Extensive experiments across diverse NVC architectures show that our method reduces the average relative bitrate error to 1.20\% and achieves up to 13.45\% bitrate savings at typical GOP sizes, outperforming existing approaches. In addition, our framework demonstrates improved robustness to content variation and bandwidth fluctuations with lower coding overhead, making it highly suitable for practical deployment.
\end{abstract}


\section{Introduction}

\label{sec:introduction}

Past years have witnessed the explosive growth of neural video compression (NVC) approaches \cite{lu2019dvc, liu2020neural, li2021deep,li2023neural, li2024neural, lu2024deep, jia2025towards}, which leverages the powerful nonlinear modeling capabilities of deep neural networks (DNNs) and end-to-end optimization to surpass traditional video coding standards in compression efficiency~\cite{bross2021overview}. Despite these breakthroughs, rate control in NVC remains underexplored, with only a few recent efforts addressing this fundamental problem~\cite{li2022rate, zhang2023neural, chen2023sparse}, although rate control is crucial for the practicality of NVC.
% only a handful of attempts have investigated the rate control of NVC~\cite{li2022rate,zhang2023neural,chen2023sparse,zhang2024learned}, even though this is vital for video codecs in real-life applications.

\begin{figure}[htbp]
\centering
\begin{minipage}[t]{0.354\linewidth}
    \centering
    \includegraphics[width=\linewidth]{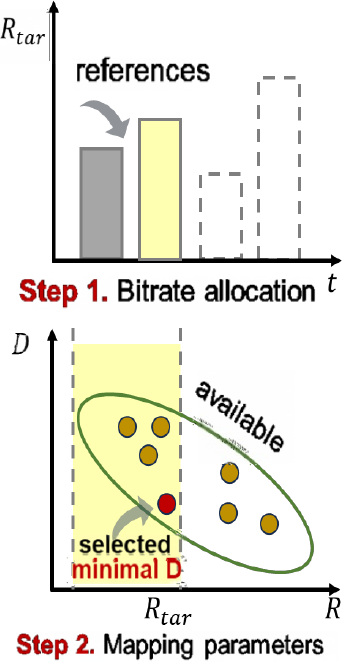}
    \small (a) Rate Control
\end{minipage}
\hfill
\begin{minipage}[t]{0.635\linewidth}
    \centering
    \includegraphics[width=\linewidth]{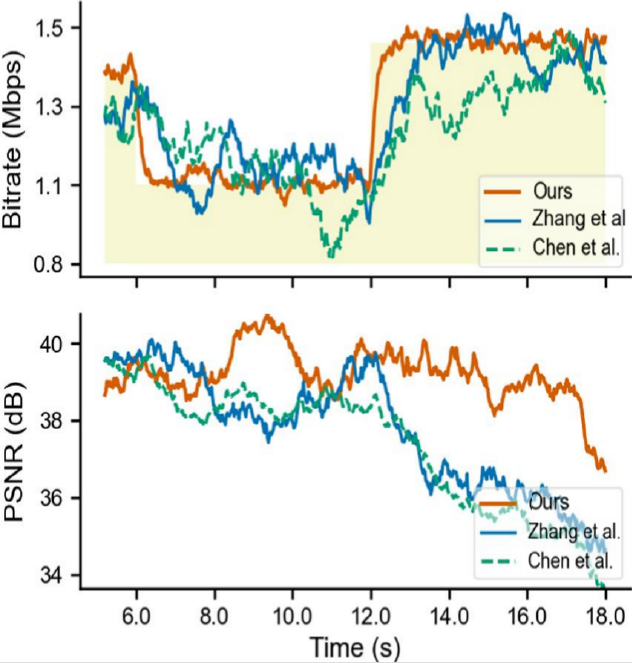}
    \small (b) Sequential Performance
\end{minipage}
\vspace{-0.2em}
\caption{Rate Control in NVC. (a) It typically involves bitrate allocation and parameter mapping.
(b) The effectiveness of rate control is reflected in its ability to accurately meet the target bitrate while incurring minimal quality degradation.}
\label{fig:fig1}
\end{figure}

Similar to traditional codecs, rate control in NVC aims to meet bitrate constraints while maximizing reconstruction quality. This typically involves dynamically adjusting coding parameters of each coding unit\footnote{Coding unit includes frame, GOP (group of pictures), etc.} (e.g., the Lagrange multiplier and/or resolution) \cite{sullivan1998rate}. Essentially, it requires learning a policy that allocates frame-level target bitrate and maps to the optimal coding parameters, as shown in Fig.~\ref{fig:fig1}(a). However, due to inter-frame dependencies, this process is affected by preceding frames and impact subsequent frames, making the global optimization problem NP-hard that computationally intractable via brute-force search.
% Due to inter‑frame dependencies, such decisions are not only influenced by previously encoded frames but also affect subsequent ones, making global optimization an NP‑hard combinatorial problem. In such case, brute‑force search is computationally and time prohibitive.

% A feasible solution is to transform the global search into a local optimization process. Specifically, previous works presumed \textit{{temporal stationarity}—within each coding unit (e.g., GOP), the rate-distortion (R\text{-}D) behavior remains consistent despite variations in contents and codec properities}—implying that simplex rule or heuristics for the entire coding unit is plausible. Under this assumption, global sequence-level optimization can be decomposed into a set of independent, tractable subproblems, reducing computational complexity to a manageable level. Various methods have been proposed to allocate bitrates and determine coding parameters within these assumed-stable units (see Fig.~\ref{fig:fig1}a).

% However, in real-world scenarios, both video content and available network bandwidth are inherently dynamic. The complex temporal reference mechanisms in NVC further violate the stationarity assumption~\cite{sheng2024predictionreferencequalityadaptation}, causing static rules and heuristic models to struggle with timely and accurate adaptation. This often leads to sub-optimal decisions, degraded R\text{-}D performance, and reduced robustness, as illustrated in Fig.~\ref{fig:fig1}b and \ref{fig:fig1}c.

A feasible solution is to approximate this global optimization problem using window‐based schemes~\cite{li2022rate, zhang2023neural, chen2023sparse}, where the target bitrate is first uniformly distributed across windows (e.g., GOPs), and then rule-based or heuristic strategies are applied within each window to allocate bitrate at the frame level.  These methods follow the philosophy of traditional codecs—leveraging content complexity to model inter-frame distortion dependencies while assuming negligible rate dependencies~\cite{hu2011rate, wang2013rate, li2020rate}. However, this assumption does not hold for NVCs. Due to their jointly optimized pixel-level, feature-based, and contextual references information, NVCs exhibit complex and tightly coupled rate and distortion dependencies~\cite{sheng2024predictionreferencequalityadaptation}. These dependencies cannot be accurately modeled using only content characteristics without incorporating information from actual encoded references. Once rate control is introduced, even minor changes in coding parameters of reference frames can lead to substantial variations in inter-frame dependencies, causing long-term shifts in the rate-distortion (R\text{–}D) behavior of both current and subsequent frames. Hence, relying solely on distortion estimations based on frame content leads to suboptimal rate allocation and coding parameters, ultimately degrading both rate accuracy and R\text{–}D performance, as shown in Fig.~\ref{fig:fig1}(b).

To tackle this, we propose a reinforcement learning (RL)-based rate control framework for NVC, formulated as a \textit{Constrained Markov Decision Process (CMDP)}~\cite{altman2021constrained}. Unlike prior schemes, our method learns a dynamic policy that jointly considers reference information, frame content and network bandwidth variations as input state. This enables the model to capture not only frame-level characteristics but also the rate and distortion dependencies induced by references. The policy directly maps states to frame-level coding parameter actions, with each action conditioned on preceding states and optimized with respect to its long-term impact. This sequential decision-making strategy enables globally-aware rate control that accounts for both frame content and sustained impacts of inter-frame dependencies in NVC.

% from environment states via exploration–exploitation, without relying on temporal stationarity. Each frame's coding parameter is recursively informed by its preceding context and optimized considering its downstream impact, enabling globally-aware, sequential decision-making. This ultimately enhances rate control accuracy and overall R\text{-}D performance.

% Although conceptually intuitive, achieving a practical balance between performance and complexity in frame-wise decision-making is non-trivial. To this end, we adopt an enhanced Actor-Critic framework~\cite{haarnoja2018soft, haarnoja2019softactorcriticalgorithmsapplications} with three key innovations: 

% 1. \textit{Neural spatiotemporal state extractor}, which captures contextual information across both spatial and temporal dimensions; 

% 2. \textit{Distributed Actor-Critic networks}, which sample and evaluate actions across a broad distribution, encouraging robust exploration and yielding more resilient policies (Fig.~\ref{fig:fig1}b);

% 3. \textit{Tempered reward design}, which modulates reward magnitudes in volatile environments to prioritize rate control precision, improving adaptability to abrupt changes (Fig.~\ref{fig:fig1}c).

Technically, we design an enhanced Actor-Critic architecture~\cite{haarnoja2018soft,haarnoja2019softactorcriticalgorithmsapplications} that integrates a neural spatiotemporal state extractor for capturing current frame contents and inter-frame dependencies, a distributed policy to support robust exploration across a diverse action space, and a tempered reward mechanism that delivers steady feedback. These components collectively enable our scheme to adapt rapidly to varying states while maintain high decision quality and efficiency as Fig.~\ref{fig:fig1}(b). Our main contributions are:
% Importantly, in contrast to RL-based rate control methods developed for traditional codecs~\cite{zhou2020rate, ho2021dual, mandhane2022muzero}, our approach is fully non-heuristic, requiring no pre-analysis or domain-specific tools, implementing a rate control scheme better suited for diverse NVCs via delicate spatiotemporal state extractor and reward mechanism. 

% Extensive experiments across various NVC architectures demonstrate that our RL-based rate control achieves state-of-the-art performance—delivering lower rate errors, greater bitrate savings across different GOP sizes, negligible time and complexity overhead, and improved robustness in unstable environments.
\begin{itemize}
    \item We conduct both theoretical and empirical analyses of the rate control problem in NVC. In contrast to traditional codecs, our study highlights the critical role of inter-frame dependencies—spanning both distortion and bitrate—in shaping rate control behavior;
    \item We propose the first CMDP-based RL framework for rate control in NVC, which integrates spatiotemporal state modeling, robust distributed exploration, and adaptive reward tempering, formulating a dynamic policy directly deciding per-frame coding parameters of NVC;
    \item We demonstrate that our method consistently outperforms prior approaches across multiple NVC architectures, achieving superior rate accuracy, bitrate savings, and robustness with minimal complexity overhead.
\end{itemize}

\section{Related Work}\label{sec:preliminaries}
\subsection{Neural Video Compression and Rate Control}\label{sec:nvc_and_rc}

NVC methods draw inspiration from traditional hybrid coding paradigms~\cite{wiegand2003overview, sullivan2012overview, bross2021overview}, designing various DNNs to implement key components such as intra-frame texture coding, inter-frame residual coding, and motion representation within an end-to-end learning framework (e.g., DVC~\cite{lu2019dvc}). These processes, including motion estimation/compensation, intra/inter prediction and residual coding, can be performed either in the original pixel domain~\cite{lu2019dvc,liu2020neural} or in a learned latent space~\cite{hu2021fvc,liu2022end}, enabling greater flexibility in modeling. The R\text{-}D trade-off is typically optimized jointly using a Lagrangian multiplier, optionally combined with a resolution scaling factor to better control bitrate and reconstruction quality~\cite{alexandre2022two}.

Following the success of the DVC series, conditional coding introduced in DCVC~\cite{li2021deep} further improved the efficiency of inter-frame feature representation and compression. Building on this, successors such as DCVC-DC~\cite{li2023neural} and DCVC-RT~\cite{jia2025towards} have demonstrated significant gains over the latest VVC standard~\cite{bross2021overview} in terms of compression performance. Despite their effectiveness, these models still operate under fixed or heuristically tuned bitrate settings during inference, limiting their adaptability to practical application scenarios.

To enable rate control in NVC, existing approaches often draw inspiration from traditional codecs~\cite{li2014lambda, liang2013novel, wang2013quadratic, li2020rate}. They divide the video sequence into windows (e.g., GOPs) and strive to fit the distortion dependencies across and within the windows, either by applying fixed R\text{–}D\text{–}$\lambda$ models~\cite{li2022rate, chen2023sparse, zhang2024learned}, or rely on empirical adjustments based on historical coding statistics~\cite{jia2025towards, yang2025neural} or heuristic pre-analysis~\cite{mandhane2022muzero, gu2024adaptiveratecontroldeep}. Then according to this preset distortion dependencies relationship, they allocate bitrate and finally map it to per-frame coding parameters. However, these methods still abide by the traditional codecs' characteristics, only consider the distortion relationships but ignore the coupled rate and distortion dependencies, which is not suitable for NVC due to its complex inter-frame references. Moreover, unlike prior overfitting methods~\cite{lu2020content, tang2024offline, chen2024group}, which are too slow, our scheme is designed to integrate a practical rate control tool into NVC.

\subsection{Reinforcement Learning}
RL focuses on learning a dynamic policy $\pi$ that selects an action $a_t$ based on the current state $s_t$ to maximize the expected cumulative return. This policy interacts with the environment by iteratively updating states and sampling actions that balance immediate rewards $r_t$ with the expected discounted sum of future rewards (discounted by factor $\gamma$), optimized via a Q-value function $Q^{\pi}(s_t, a_t)$. In this framework, the global optimization problem is recast as a sequential decision‐making process over a temporal horizon:
\begin{align}
\label{eq:q-function}
    Q^{\pi}(s_t,a_t) \!=\! r(s_t, a_t) \!+\! \gamma \mathbb{E}_{s_{t+1}, a_{t+1}}[Q^{\pi}(s_{t+1},a_{t+1})].
\end{align}

At each timestep $t$, future outcomes are uncertain due to the non-deterministic state transition $\mathcal{P}(s_{t+1}|s_t, a_t)$. This uncertainty is analogous to rate control in NVC, where future content, network bandwidth, and codec behavior may change unpredictably. Fundamentally, this scenario reflects the exploration-exploitation dilemma central to RL. In this work, we propose an enhanced Actor-Critic framework~\cite{konda1999actor, haarnoja2018soft, haarnoja2019softactorcriticalgorithmsapplications} that improves state representation, action sampling, and reward shaping, yielding a more practical and effective trade-off between performance and adaptability.

While RL techniques have been explored in rate control for traditional codecs, they either build on traditional rule‐based tools by using RL to explore better rules~\cite{9088297, 9418757, gadot2025rl}, or rely on heuristic search methods that require multiple pre‐codings~\cite{mao2020neural, mandhane2022muzero}. In essence, none of them explicitly account for the coupled inter‐frame rate and distortion dependencies in NVC, causing suboptimal performance. Moreover, the lack of robust existing tools and dependence of multiple pre‐codings further restrict their applicability to NVC.

\section{Rate Control for NVC}\label{sec:3}
In this section, we first provide a re-analysis of the rate control problem in NVC, followed by an in-depth analysis of how rate and distortion coupled dependencies impact rate control.

\subsection{Problem Formulation} \label{sec:sec3_1}

Rate control for a video sequence $\mathcal{X} = \{x_1, x_2, \ldots, x_N\}$ of length $N$ can be formulated as a constrained optimization problem. Given a target bitrate $R_{tar}$ imposed at a specific level (e.g., sequence or GOP level), the goal is to determine an optimal set of frame-wise coding parameters $\Pi^{(\mathcal{X})} = \{\boldsymbol{a}_1, \boldsymbol{a}_2, \ldots, \boldsymbol{a}_N\}$ that minimizes the total distortion: 
\begin{align}
    \Pi^{(x_t)} = \arg\min_{\Pi^{(\mathcal X)}} \sum_{t=1}^N D_t , \ \text{s.t.} \ \frac{1}{N} \sum_{t=1}^N R_t \leq R_{tar},
\label{eq:rate_control}
\end{align}
This constrained problem can be equivalently reformulated in an unconstrained form by introducing a global Lagrangian multiplier $\Lambda$ that balances distortion minimization and rate constraint satisfaction:
\begin{align}
\Pi^{(x_t)}=\arg\min_{\Pi^{(\mathcal X)}} \sum_{t=1}^N D_t + \Lambda (\frac{1}{N}\sum_{t=1}^N R_t - R_{tar}).
\label{eq:unconstrained_rc}
\end{align}

Taking the derivative of Eq.~\eqref{eq:unconstrained_rc} with respect to each $\mathbf{a}_t$ yields the necessary condition for the optimal parameters set:
\begin{align}
\frac{\partial \sum_{t=1}^{N} D_t}{\partial \mathbf{a}_t}
+ \frac{\Lambda}{N} \frac{\partial \sum_{t=1}^{N} R_t}{\partial \mathbf{a}_t} = 0,
\quad t = 1, 2, \cdots, N.
\label{eq:4}
\end{align}
According to R\text{-}D theory, the R\text{-}D function is a convex, and its slope at each point is given by $\lambda_t={\partial D_t} / {\partial R_t}$. In traditional video codecs, it is typically assumed that inter-frame bitrate dependencies are negligible, and distortion dependencies are approximated using fixed rules or heuristics~\cite{hu2011rate, wang2013rate, li2020rate}. Under these assumptions, Eq.~\eqref{eq:4} can be transformed into the following optimality criterion:
\begin{align}
\lambda_t = \frac{\Lambda}{N \cdot\frac{\partial \sum_{i=t}^N D_t}{\partial D_t}}=\omega_t \cdot \Lambda, t = 1,2,\cdots,N.
\label{eq:trad_action_decision}
\end{align}

However, these assumptions do not hold in jointly-trained, non-linear NVCs. Their complex pixel-level, feature-based, and contextual dependencies not only introduce strong inter-frame distortion dependencies but also lead to tightly coupled bitrate dependencies. For example, temporal context modeling directly affects the estimated probability distributions of latent features, thereby influencing the actual bitrate. As a result, Eq.~\eqref{eq:4} must be revised to capture these interactions:
\begin{align}
\sum_{i=t}^N\big(\frac{\Lambda}{N}\frac{\partial R_i}{\partial \mathbf{a}_t} - \frac{\partial D_i}{\partial \mathbf{a}_t}\big) = \sum_{i=t}^N\Big(\big(\frac{\Lambda}{N}-\lambda_i\big) \frac{\partial R_i}{\partial \mathbf{a}_t}\Big) = 0,
\label{eq:action_decision}
\end{align}
which implies that the optimal coding parameter $\mathbf a_t$ must not only reflect the frame's R\text{-}D behavior (i.e., $\lambda_t$) but also consider propagated rate and distortion impact on future frames. 

% \begin{figure}[htbp]
% \centering
% \begin{minipage}[t]{0.495\linewidth}
%     \centering
%     \includegraphics[width=\linewidth]{figures/subplot_1.pdf}
%     \small (a) Original
% \end{minipage}
% \hfill
% \begin{minipage}[t]{0.495\linewidth}
%     \centering
%     \includegraphics[width=\linewidth]{figures/subplot_2.pdf}
%     \small (b) w/ Manual Modeling, \citet{chen2023sparse}
% \end{minipage}

% \begin{minipage}[t]{0.495\linewidth}
%     \centering
%     \includegraphics[width=\linewidth]{figures/subplot_3.pdf}
%     \small (c) w/ Heuristic Methods, \citet{jia2025towards}
% \end{minipage}
% \hfill
% \begin{minipage}[t]{0.495\linewidth}
%     \centering
%     \includegraphics[width=\linewidth]{figures/subplot_4.pdf}
%     \small (d) w/ NN-based Prediction, \citet{zhang2023neural}
% \end{minipage}
% \caption{{\bf R-D characteristic of Frame 10 with different rate control methods. }The orange $\times$ markers are the actual R-D operation points with different $\lambda$, while the blue line is a fitted hyperbolic function. Target bitrate constraint is marked by a yellow rectangle, finally the yellow $\circ$ and gold $\star$ are the optimal R-D points decision with/without bitrate constraint. (Tested on the \textit{Beauty} sequence with DCVC-RT).}
% \label{fig:reference_rd}
% \end{figure}

\subsection{The Impact of Inter-frame Dependencies}~\label{sec:inter_NVC}
% This subsection analyzes how integrating rate control modules impacts frame-level R\text{-}D behavior in NVC, emphasizing the need for adaptive modeling to guide per-frame bitrate allocation and coding parameter selection.

Typically, a pretrained NVC is optimized to adapt to a specific R\text{-}D dependency under a fixed global $\Lambda$ constraint.
However, once rate control is introduced, frame-wise variations in $\lambda$ induce new rate and distortion dependencies that diverge from those learned during pretraining.
% Existing schemes only strive to gain some preset weights $\omega_t$ in Eq.~\eqref{eq:trad_action_decision} based on frames contents characteristics before actually encoding, to fit the inter-frame distortion dependencies, but ignore the coupled impact of rate and distortion dependencies. Finally they often lead to suboptimal rate allocation and coding parameters, ultimately degrading rate accuracy and R-D performance.

% (as well as codecs with hierarchical quality~\citet{li2023neural, jia2025towards}, which integrates fixed per-frame $\lambda$ offset patterns, but the final loss function still employs a single $\Lambda$ value in each training sample). 

\begin{figure}[tbp]
\centering
\includegraphics[width=0.76\linewidth]{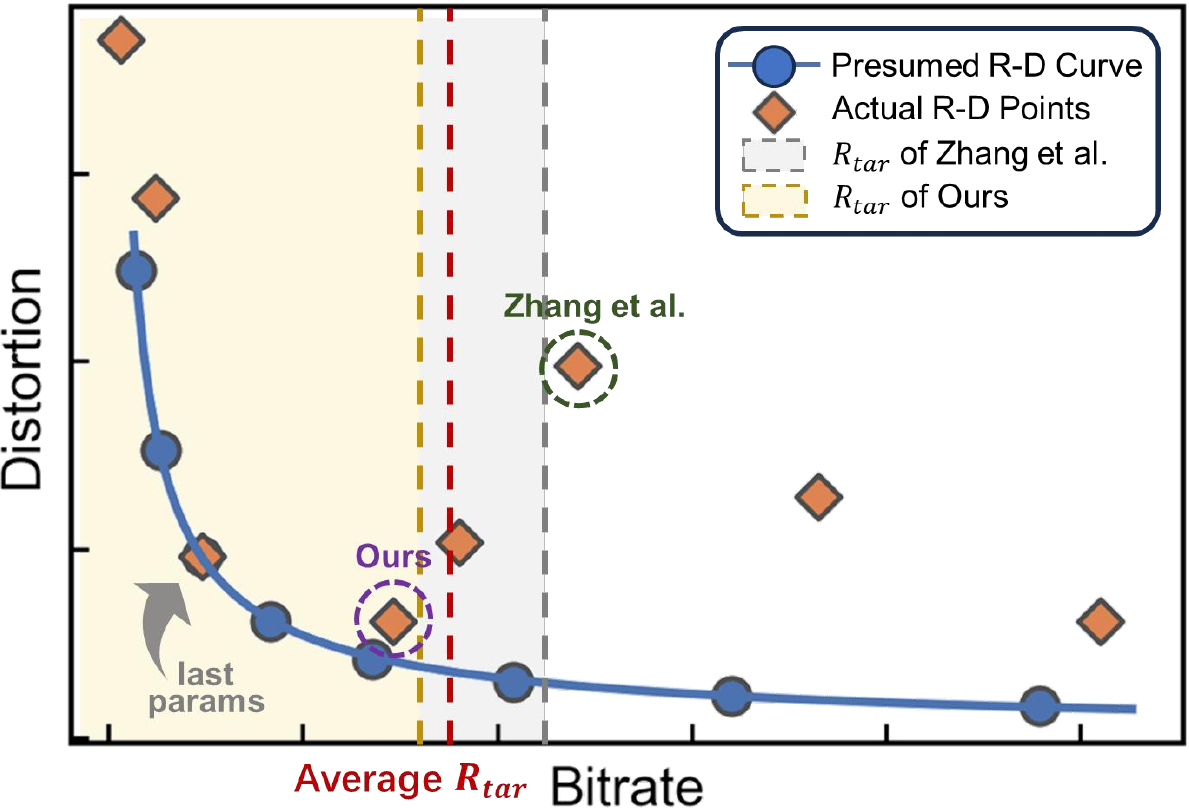}
\vspace{-0.2em}
\caption{Rate Control Process for the 25-th Frame on \textit{BasketballDrive}. Ignoring inter-frame rate dependencies leads to improper bitrate allocation. As a result, the coding decision still follows the pretrained R\text{–}D curve (blue), producing suboptimal parameters (green “Zhang et al.”).}
\label{fig:fig2}
\end{figure}

To further investigate the impact of inter-frame dependencies-formulated in Eq.~\eqref{eq:action_decision}-on rate control in NVC, we conduct a toy experiment on the \textit{BasketballDrive} sequence using the DCVC‑DC codec. Following the state-of-the-art rate control method for NVC~\cite{zhang2023neural}, we set the target bitrate for the entire sequence to match the average bitrate achieved by standard fixed-QP coding with $QP=32$, where the quantization parameter (QP) corresponds one-to-one with $\lambda$. Fig.~\ref{fig:fig2} illustrates the rate control behavior for the 25-th frame. The blue solid line represents the assumed R\text{–}D curve for the frame under the pretrained model, typically a smooth hyperbolic function that does not account for reference frame impacts. However, in the rate controlled setting, the 24-th reference frame may be encoded with a mismatched QP (e.g., QP=12), leading to altered inter-frame dependencies. This discrepancy shifts the R\text{–}D behavior of the 25th frame, causing it to deviate from the original trajectory—as shown by the scattered orange points.

\begin{figure*}[htbp]
\centering
\includegraphics[width=0.88\linewidth]{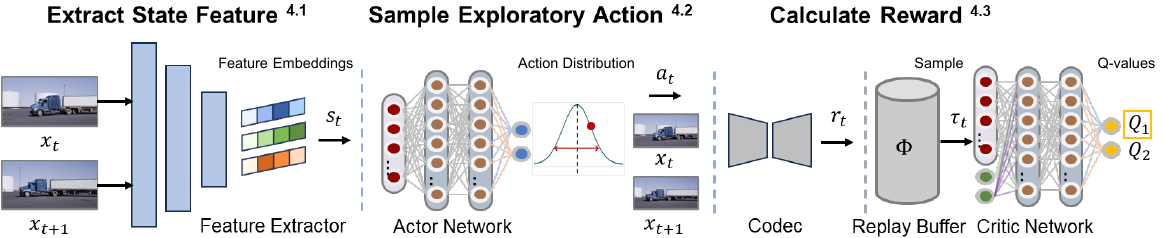}
\vspace{-0.2em}
\caption{Training Pipeline of the Proposed Actor-Critic Network. The entire process consists of three main components, from state to action and then to reward, corresponding to Sec.~\ref{sec:state_modeling}, \ref{sec:action_decision} and \ref{sec:reward_shaping}, in sequence.}
\label{fig:training_pipeline}
\end{figure*}

Since prior methods do not account for reference frame information produced during actual encoding, they cannot capture this shift. As a result, they allocate an inappropriate bitrate and select coding parameters that minimize distortion based on a static R\text{–}D assumption—corresponding to the 6th point on the blue curve. However, due to the reference QP mismatch, the actual operating point drifts to the 6th orange point (denoted as “Zhang et al.”), leading to suboptimal performance and potential bitrate overshoot. In contrast, our proposed framework more accurately models the true R\text{–}D behavior, enabling more effective bitrate allocation. Consequently, the selected coding parameters better align with the target average bitrate while achieving lower distortion—corresponding to the 4th point (denoted as “Ours”).

These findings reveal a critical limitation in existing NVC rate control methods, echoing the observations in Sec.~\ref{sec:sec3_1}: both frame-level R\text{–}D characteristics and inter-frame rate and distortion dependencies must be considered. To this end, we propose an RL-based rate control framework that explicitly models frame content and reference-induced dependencies, and directly determines per-frame coding parameters by maximizing the expected cumulative reward. This approach inherently aligns with the formulation in Eq.~\eqref{eq:unconstrained_rc}, with further technical details provided in Sec.~\ref{sec:4}.

\section{Reinforced Rate Control Framework}\label{sec:4}

As shown in Eq.~\eqref{eq:action_decision}, the core challenge lies in accurately modeling frame-wise states while exploring each action's long-term impact on future frames. To tackle this, we design an enhanced Actor–Critic framework (illustrated in Fig.~\ref{fig:training_pipeline}), with the following key innovations.

\subsection{State Modeling} \label{sec:state_modeling}
In RL, an informative and compact state representation is crucial for both effective policy learning and generalization~\cite{li2006towards, mnih2015human}. According to Eq.~\eqref{eq:action_decision}, the state in rate control should encapsulate both historical encoded references and current frames. Unlike prior schemes that rely on handcrafted features~\cite{chen2018reinforcement, zhou2020rate}, we learn this representation end-to-end using a neural embedding network conditioned on current frame, references, and auxiliary information.

Specifically, the current frame $x_t$ and reference frame $x_{t-1}$ are concatenated and passed through a cascaded residual network for spatial-temporal feature extraction. Additionally, intermediate features of $x_{t-1}$-extracted by the codec at multiple resolutions-are fused to enhance the temporal context. The resulting embeddings are further refined using several convolutional layers and average pooling operations. To supplement visual context, auxiliary information such as the target bitrate and previously selected coding parameters is normalized, expanded, and embedded through fully connected layers. The combined visual and auxiliary embeddings form a comprehensive, learnable state representation. 

Compared to prior methods, our learned state embedding flexibly incorporates dynamic references and frames' characteristics, enabling more accurate and adaptive modeling of R\text{–}D behavior for frame-wise decision-making. Detailed architecture and setups are included in the Appendix.

\subsection{Action Decision} \label{sec:action_decision}

The action in our RL framework is defined as a pair of continuous coding parameters $\{\lambda_t, m_t\}$, where $\lambda_t \in \left[\lambda_{min}, \lambda_{max}\right]$ denotes the Lagrange multiplier controlling the $R\text{-}D$ behavior, and $m_t \in \left[ 0.5, 1.0 \right]$ is down-sampling factor that adjusts the spatial resolution of both the current and reference frames. 

Given the continuous and high-dimensional nature of the action space, we model the policy $\pi_{\phi}$ as a Gaussian distribution, where both the mean and variance are predicted by the Actor network $\phi$. This probabilistic formulation allows for exploration beyond deterministic actions used in prior rate control strategies and improves adaptability across diverse states. To further encourage exploration, we incorporate policy entropy regularization~\cite{haarnoja2018soft} and optimize the Actor using the policy gradient:
\begin{equation}
\label{loss_for_actor}
    J_\pi(\phi) \! =\! \mathbb{E}_{s_t \sim \mathcal{S}, a_t \sim \pi_\phi} \left[ \epsilon \log \pi_\phi(a_t | s_t)\! -\! Q_\theta(s_t, a_t) \right],
\end{equation}

During inference, we adopt a greedy strategy by selecting the action with the highest likelihood. If resolution scaling ($m_t < 1.0$) occurs, we resample the reference frame to match the current resolution for consistent inter-frame prediction, then bicubically upsample the output to the original resolution. This joint $\lambda$\text{–}$m$ policy not only provides precise rate control but also reduces computational cost by enabling lower-resolution processing when appropriate.

% For inference, a greedy strategy selects the action with the highest probability. Additionally, $m_t$ may induce resolution differences. In such cases: The reference frame is resampled to match the current frame's resolution, ensuring proper inter-frame prediction. Finally, the reconstructed frame undergoes bicubic upsampling to restore its original resolution. This joint $\lambda \text{-}m$ action policy achieves precise rate control while accelerating the coding process by reducing computational complexity due to lower resolutions. 

\begin{table*}[t]
    \centering
    \renewcommand{\arraystretch}{0.93}
    \setlength{\tabcolsep}{9.5pt}
    \large
    \resizebox{0.93\textwidth}{!}{%
    \begin{tabular}{cllccccccc}
        \toprule
        \multirow{2}{*}{GOP} & \multirow{2}{*}{Codec} & \multirow{2}{*}{Method} & \multicolumn{7}{c}{Dataset} \\
        \cmidrule(lr){4-10}
         &  &  & UVG & MCL-JCV & HEVC B & HEVC C & HEVC D & HEVC E & Avg. \\
        \midrule
        % GOP = 32
        \multirow{12}{*}{32} 
        & \multirow{3}{*}{DVC} 
          & Zhang et al. & --- & --- & --- & --- & --- & --- & --- \\
        &  & Chen et al.  & 1.64 / -14.57 & 2.11 / -12.34  & 1.69 / -15.01 & 1.45 / -14.72 & 1.38 / -13.89 & 1.65 / -20.23 & 1.65 / -15.13 \\
        &  & Ours         & \textbf{1.32 / -16.97} & \textbf{1.82 / -17.11}  & \textbf{1.23 / -16.64} & \textbf{1.12 / -16.28} & \textbf{1.08 / -15.97} & \textbf{1.23 / -21.43} & \textbf{1.30 / -17.40} \\
        \cmidrule(lr){2-10}
        & \multirow{3}{*}{DCVC} 
          & Zhang et al. & --- & --- & --- & --- & --- & --- & --- \\
        &  & Chen et al.  & 1.85 / -14.28 & 2.03 / -10.88 & 1.96 / -12.23 & 1.73 / -9.47  & 1.81 / -10.98 & 1.18 / -15.25 & 1.76 / -12.18 \\
        &  & Ours         & \textbf{1.80 / -18.24} & \textbf{1.79 / -17.11} & \textbf{1.15 / -14.83} & \textbf{1.65 / -14.98} & \textbf{1.51 / -15.03} & \textbf{0.99 / -18.76} & \textbf{1.48 / -16.49} \\
        \cmidrule(lr){2-10}
        & \multirow{3}{*}{DCVC-DC} 
          & Zhang et al. & --- & --- & --- & --- & --- & --- & --- \\
        &  & Chen et al.  & 1.66 / -10.33 & 1.83 / -9.92  & 1.71 / -11.02 & 1.55 / -9.67  & 1.60 / -9.84  & 1.28 / -13.00 & 1.61 / -10.63 \\
        &  & Ours         & \textbf{1.45 / -13.84} & \textbf{1.57 / -12.51} & \textbf{0.98 / -14.82} & \textbf{0.97 / -13.03} & \textbf{0.93 / -12.98} & \textbf{0.85 / -16.70} & \textbf{1.13 / -13.98} \\
        \cmidrule(lr){2-10}
        & \multirow{3}{*}{DCVC-RT} 
          & Zhang et al. & --- & --- & --- & --- & --- & --- & --- \\
        &  & Chen et al.  & 1.49 / -5.12 & 1.71 / \textbf{-5.33} & 1.44 / -5.26 & 1.35 / -4.08 & 1.22 / -4.10 & 1.50 / -4.98 & 1.45 / -4.81 \\
        &  & Ours         & \textbf{1.18 / -5.84} & \textbf{1.27} / -5.31 & \textbf{1.16 / -6.00} & \textbf{1.09 / -4.79} & \textbf{1.23 / -4.86} & \textbf{0.96 / -6.17} & \textbf{1.15 / -5.50} \\
        \midrule
        % GOP = 100
        \multirow{12}{*}{100} 
        & \multirow{3}{*}{DVC} 
          & Zhang et al. & 2.82 / -11.61 & 2.79 / -8.78  & 1.35 / -10.99 & 1.18 / -10.63 & 1.91 / -12.17 & \textbf{1.11} / -18.28 & 1.86 / -12.08 \\
        &  & Chen et al.  & 1.73 / -16.11 & 2.16 / -13.08 & 1.68 / -15.33 & 1.66 / -16.02 & 1.50 / -14.54 & 1.79 / \textbf{-20.15} & 1.75 / -15.87 \\
        &  & Ours         & \textbf{1.30 / -17.04} & \textbf{1.88 / -17.28} & \textbf{1.20 / -16.63} & \textbf{1.01 / -16.55} & \textbf{1.05 / -16.14} & \textbf{1.21} / -20.05 & \textbf{1.28 / -17.28} \\
        \cmidrule(lr){2-10}
        & \multirow{3}{*}{DCVC} 
          & Zhang et al. & 2.80 / -7.34 & 2.95 / -5.68  & 2.32 / -5.88  & 1.94 / -4.42  & 2.11 / -3.80  & 1.33 / -9.24 & 2.24 / -6.06 \\
        &  & Chen et al.  & 1.81 / -14.33 & 2.02 / -11.02 & 2.01 / -12.60 & 1.64 / -9.93 & 1.80 / -10.89 & 1.22 / -15.31 & 1.75 / -12.35 \\
        &  & Ours         & \textbf{1.77 / -18.19} & \textbf{1.79 / -17.30} & \textbf{1.18 / -14.88} & \textbf{1.58 / -14.91} & \textbf{1.47 / -15.22} & \textbf{1.02 / -18.77} & \textbf{1.47 / -16.55} \\
        \cmidrule(lr){2-10}
        & \multirow{3}{*}{DCVC-DC} 
          & Zhang et al. & 2.25 / -6.50  & 2.08 / -5.24  & 1.74 / -4.33  & 1.62 / -4.20  & 2.03 / -3.99  & 1.37 / -8.17 & 1.85 / -5.41 \\
        &  & Chen et al.  & 1.69 / -9.99 & 1.81 / -9.88 & 1.72 / -11.13 & 1.54 / -9.41 & 1.66 / -9.89 & 1.35 / -12.19 & 1.62 / -10.42 \\
        &  & Ours         & \textbf{1.40 / -13.64} & \textbf{1.52 / -12.55} & \textbf{0.95 / -14.93} & \textbf{0.92 / -12.88} & \textbf{0.91 / -13.05} & \textbf{0.83 / -16.52} & \textbf{1.09 / -13.93} \\
        \cmidrule(lr){2-10}
        & \multirow{3}{*}{DCVC-RT} 
          & Zhang et al. & --- & --- & --- & --- & --- & --- & --- \\
        &  & Chen et al.  & 1.33 / -5.67 & 1.50 / -5.11 & 1.43 / \textbf{-6.31} & 1.25 / -4.40 & 1.36 / -4.76 & 1.02 / -6.06 & 1.32 / -5.39 \\
        &  & Ours         & \textbf{1.02 / -6.37} & \textbf{1.25 / -6.15} & \textbf{0.91} / -6.27 & \textbf{0.87 / -5.13} & \textbf{0.96 / -5.09} & \textbf{0.85 / -7.14} & \textbf{0.98 / -6.03} \\
        \bottomrule
    \end{tabular}%
    }
    \vspace{-0.2em}
    \caption{Performance comparison across datasets ($\Delta R\downarrow$ / BD-Rate (\%)$\downarrow$) with average performance. Bold values indicate the best performance. {(Due to the lack of results at the GOP size of 32 in~\citet{zhang2023neural} and with DCVC-RT, where we have provisionally excluded comparisons with~\citet{zhang2023neural} to prevent potential discrepancies.)}}    \label{tab:overall_performance}
\end{table*}

\subsection{Reward Shaping} \label{sec:reward_shaping}
% Rewards serve as feedback to evaluate actions' values. However, due to objectives such as rate control accuracy emerge only after coding an entire sequence, the rewards are inherently sparse. While prior methods that derive per-frame rate allocation—using off‐the‐shelf tools~\cite{zhou2020rate, ho2021dual}, heuristic~\cite{mandhane2022muzero}—to provide instant reward, but no universal scheme exists in NVC.

Rewards serve as the learning signal to evaluate the quality of selected actions. However, in rate control tasks, meaningful metrics such as total distortion or bitrate deviation are only available after coding an entire sequence, making rewards inherently sparse. While previous methods attempt to define intermediate rewards using off-the-shelf tools~\cite{zhou2020rate, ho2021dual}, heuristics~\cite{mandhane2022muzero}, or fixed allocation rules, yet no general solution exists for NVC.

% This rises a core problem about trade‐off between exploration and exploitation: overly dense, per‐frame rewards can accelerate convergence but stifle discovery of superior strategies; overly sparse rewards leave intermediate frames without guidance~\cite{bellemare2016unifying, saunders2017trial}. Hence, we reshape the reward as a weighted inner product of distortion and rate error terms in Eq.~\eqref{eq:action_decision}.
This poses a fundamental exploration and exploitation dilemma: overly dense, per-frame rewards may accelerate convergence but discourage discovery of better global strategies; overly sparse rewards leave instant frames without guidance~\cite{bellemare2016unifying, saunders2017trial}. To strike a balance, we reshape the reward as a weighted inner product of distortion and rate deviation terms:
\begin{align}
    r_t = -\mathbf w_t^{\!\top}\,\mathbf f_t
    ,\quad
    \mathbf f_t =
    \begin{pmatrix}
    D_t \\
    \dfrac{\lvert R_{\rm rem}\rvert}{R_{\rm tar}}
    \end{pmatrix}
    ,\quad
    \mathbf w_t =
    \begin{pmatrix}
    \delta_t \\[6.5pt]
    \eta_t
    \end{pmatrix}.
    \label{eq:reward1}
\end{align}
where $R_{rem}$ is the remaining bitrate budget. The weight vector $\mathbf w_t=(\delta, \eta)^\top$ balances distortion and rate accuracy, and is periodically updated every $\mathcal{K}$ training steps using validation feedback. For the final frame, a large $\eta_t$ is applied to enforce strict rate control. To prevent overspending the bitrate budget, over-allocation is penalized accordingly.

Furthermore, We adopt a twin‐Critic architecture where two independent Q-values are estimated, and their minimum is used to mitigate overestimation bias~\cite{hasselt2010double}. In addition, we model the full return distribution instead of a scalar expected value, enhancing robustness in reward estimation~\cite{bellemare2017distributional}.

% Unlike prior methods, our proposed RL‐based scheme dynamically shapes rewards without relying on fixed rules or heuristics, enhancing the policy effectiveness and robustness. Details of reward shaping can be found in the Appendix.

Unlike previous schemes that rely on fixed allocation rules or handcrafted heuristics, our proposed RL-based framework adaptively shapes rewards, improving the learned policy’s generalization and effectiveness. Implementation details and related hyperparameters are available in the Appendix.

\section{Experiments}\label{sec:experiments}
\subsection{Experimental Setup} \label{sec:section4-a}
\textbf{Base Codecs}: We perform evaluation on four representative NVCs: DVC \cite{lu2019dvc}, DCVC \cite{li2021deep}, DCVC-DC \cite{li2023neural} and the latest DCVC-RT \cite{jia2025towards}.
Since DVC and DCVC are fixed-rate, we extend their highest bitrate pretrained models to support variable-rate coding according to \citet{duan2023qarv}, with their original training setups. For DCVC-DC and DCVC-RT, we directly adopt their pre-trained variable-rate model. To achieve flexible rate control, we set $m_t \in \left[0.5, 1.0\right]$, $\lambda_t \in \left[256, 2048 \right]$ for DVC and DCVC, $\lambda_t \in \left[85, 840 \right]$ for DCVC-DC, and $\lambda_t \in \left[1, 768 \right]$ for DCVC-RT according to their original setup. Details can be found in the Appendix.

\textbf{Datasets}: For training our rate control module (parameters of codecs remain fixed), we construct a mixed dataset combining BVI-DVC \cite{ma2021bvi} and Vimeo sequences \cite{xue2019video}. For validation, we use the USTC-TD dataset \cite{li2024ustc}. For evaluation, we follow standard benchmarks, selecting UVG dataset \cite{mercat2020uvg}, MCL-JCV dataset \cite{wang2016mcl}, and HEVC Class B to E sequences \cite{bossen2013common}.

\textbf{Implementation Details}: All experiments are implemented using the PyTorch framework on an NVIDIA RTX 3090 GPU. {To enhance sample efficiency during RL training, we adopt a replay buffer of length 200 for offline updates, sampling 32 trajectories per iteration.} Initially, all networks are pretrained for 50 epochs using 4-frame sequences. The feature extractor is then fixed, and training continues for an additional 250 epochs with 32-frame sequences. Further training strategies and details are provided in the Appendix.

\textbf{RC Benchmarks:} 
% We compare our approach against two state-of-the-art methods: (i) ~\citet{chen2023sparse} models (R\text{-}$\lambda$\text{-}m, D\text{-}$\lambda$\text{-}m) relationships using a hyperbolic function with iterative updates. (ii) ~\citet{zhang2023neural} uses a neural network to predict the rate allocation and R\text{-}$\lambda$ mapping. {For a fair comparison, we use the same test conditions (e.g. all frames are evaluated, and GOP size is 32 or 100 with LDP configuration).} We emphasize that our method offers a universal, plug‑and‑play module for existing NVC systems, while some other schemes are either not generalizable or address different concerns. Discussion and comparisons with these works can be found in the Appendix.
We compare our approach with two state-of-the-art methods: (i) ~\citet{chen2023sparse}, which models the R\text{–}$\lambda$\text{–}m and D\text{–}$\lambda$\text{–}m relationships using a hyperbolic function with iterative updates; and (ii) ~\citet{zhang2023neural}, which employs a neural network to predict rate allocation and the R\text{–}$\lambda$ mapping. For a fair comparison, we adopt the same test conditions across all methods—for example, evaluating all frames and using a GOP size of 32 or 100 under an LDP configuration. Notably, our approach is intended to explore a general, plug‑and‑play rate‑control method for NVC; schemes designed for traditional codecs, or those tailored to specific architectures or other objectives, are not within the scope of comparisons. Details can be found in the Appendix.

\begin{figure}[t]
\centering
\includegraphics[width=1.0\linewidth]{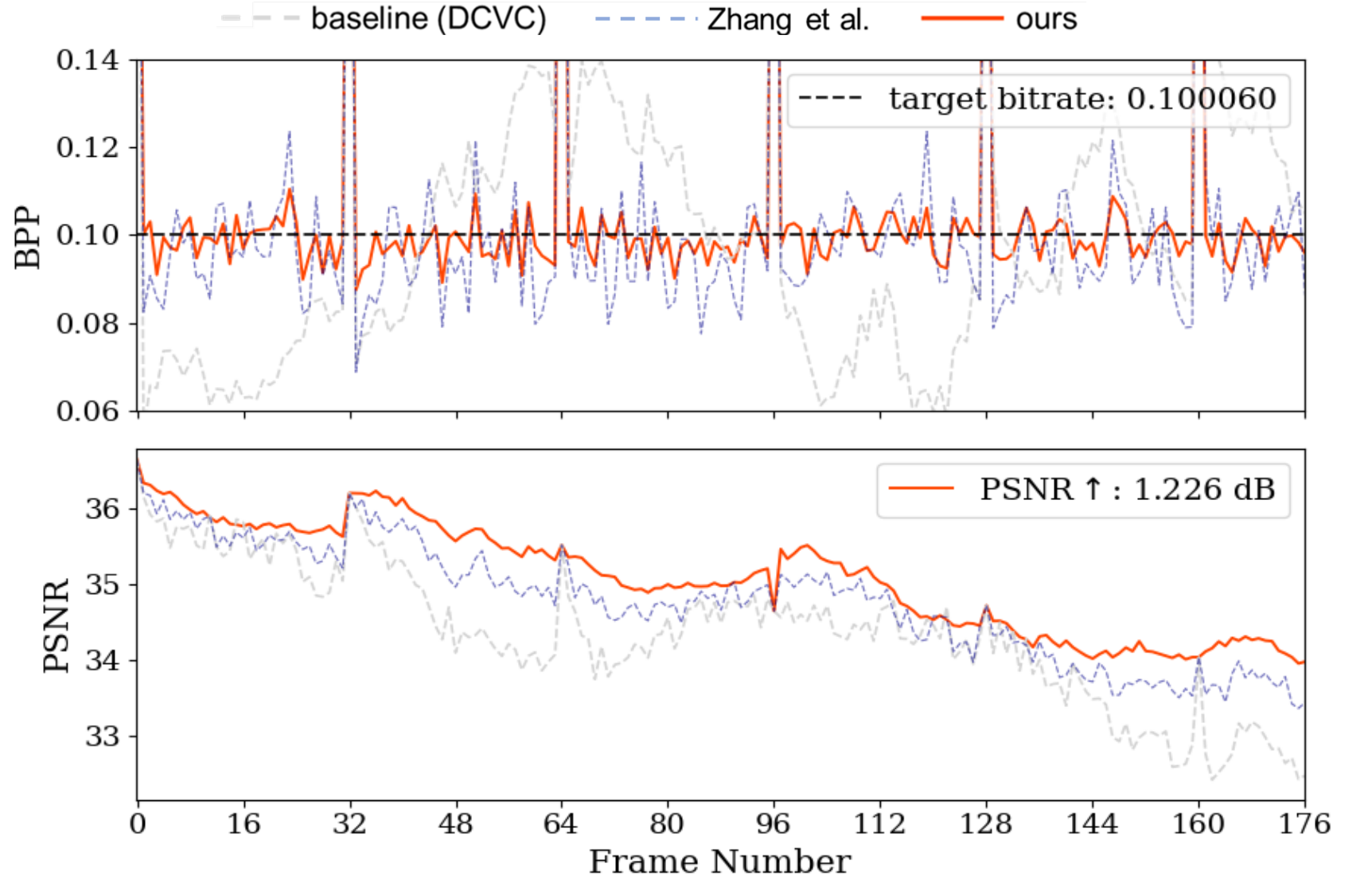}
\vspace{-1.5em}
\caption{Frame-Level Rate Control. Evaluated with a unified target bitrate \textit{0.10006} BPP on \textit{BasketballDrive} sequence.}
\label{fig:rate_control1}
\end{figure}

\subsection{Experimental Results}
\textbf{Performance Analysis}: 
As shown in Table \ref{tab:overall_performance}, our RL-based method demonstrates superior performance in both rate control accuracy and $R\text{-}D$ performance across various GOP sizes. For a GOP size of 32, it achieves lower rate errors across almost all evaluated codecs and datasets, along with higher BD-Rate gains. Even when applied to NVC frameworks with complex inter-frame dependencies, such as DCVC-DC, our method achieves a rate error of just 1.13\% and a bitrate saving of 13.98\%, verifying the effectiveness of the proposed RL-based scheme. In DCVC‑RT-whose native models are trained with hierarchical quality on long sequences and thus inherently perform implicit rate allocation to enhance R\text{-}D performance-the integration of an additional rate control module brings only marginal improvements. Nevertheless, our method maintains a low rate error of just 1.15\%, indicating strong robustness. For a longer GOP size of 100, our approach continues to demonstrate clear advantages. Notably, as the underlying codec improves (e.g., from DVC to DCVC-RT), the BD-Rate gains achieved by the method of \citet{zhang2023neural} diminish significantly—shrinking to just 5.41\% on DCVC-DC—whereas our approach maintains a stable 13.93\% gain, highlighting its superior generalization capability.

To further evaluate the effectiveness of our RL-based scheme, we visualize per-frame performance and compare it with the method of \citet{zhang2023neural} on DCVC. Specifically, (i) \textit{Fixed Target Bitrate:} We encode the entire sequence under a constant bitrate constrain to simulate stable network conditions. As shown in Fig.~\ref{fig:rate_control1}, our method exhibits significantly lower rate fluctuation and reduced quality degradation across frames. (ii) \textit{Dynamic Network Bandwidth:} In real-world streaming scenarios, network bandwidth varies over short time intervals, necessitating adaptive rate allocation. To simulate this, we use real-life bandwidth traces from \citet{fcc_raw_data} to set target bitrate. As illustrated in Fig.~\ref{fig:fig1}(b), our method adapts more effectively to bandwidth fluctuations, achieving smoother bitrate transitions and more stable quality. In contrast, other methods struggle to cope with such variations. This finding also supports our discussion in Sec.\ref{sec:inter_NVC}—the necessity to consider both frames contents characteristics and inter-frame rate and distortion dependencies in rate control of NVC.

% \begin{figure}[t]
% \centering
% \includegraphics[width=0.95\linewidth]{figures/rate_control2.pdf}
% \caption{{\bf Per-frame rate control effect with a varying target bitrate \textit{0.0163 $\rightarrow$ 0.0173 $\rightarrow$ 0.0166} BPP on \textit{Johnny} sequence.} The rate changes every 0.5 seconds, i.e., 12 frames.}
% \label{fig:rate_control2}
% \end{figure}

\begin{figure}[t]
\centering
\includegraphics[width=1.0\linewidth]{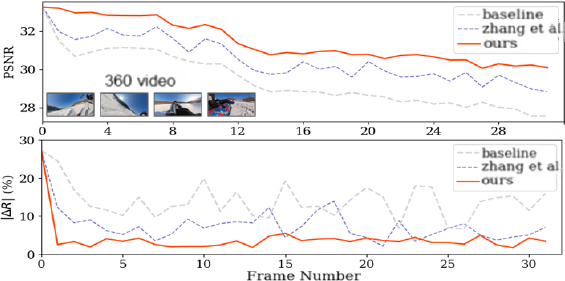}
\vspace{-1.5em}
\caption{Performance Comparison over 360$^{\circ}$ Video. The test sequence is downloaded from \url{https://www.youtube.com/watch?v=4T8yFnHaJtc}, with the results of quality degradation above and rate fluctuation below.}
\label{fig:360_video}
\end{figure}

\textbf{Generalization Analysis}: 
An effective rate control scheme must generalize well to diverse video contents and motion patterns. To assess this, we evaluate our method on an unseen 360-degree video sequence, which exhibits substantially greater content variability and motion dynamics than those seen during training. As shown in Fig.~\ref{fig:360_video}, the naive baseline suffers from significant quality decline and pronounced rate fluctuations. While \citet{zhang2023neural} partially alleviates the rate fluctuation, it still exhibits a PSNR drop exceeding 4.4 dB and a rate bias of approximately 7.6\%. In contrast, our method demonstrates a smaller PSNR degradation and maintains a lower rate bias of ~3.9\%. These results underscore the superior generalization capability of our approach. This improvement is attributed to our balanced exploration–exploitation strategy—enabled by random action sampling and stable reward feedback—which facilitates a more adaptive and robust policy when encountering volatile content distributions. Additional generalization comparisons with other methods are provided in the Appendix.

% We therefore evaluate this on an unseen 360-degree video with far greater content variation and motion diversity than the training set. As Fig.~\ref{fig:360_video}, naive baseline suffers significant error propagation and rate fluctuation. \citet{zhang2023neural} partially mitigates rate fluctuation, but still suffers from a $>4.4$ dB PSNR drop and a $~7.6\%$ rate bias. In contrast, our method achieves a smaller PSNR drop and maintains a $~3.9\%$ rate bias. These results highlight the stronger generalization of our approach, which benefits from its fine exploration-exploitation tradeoff via random actions sampling and steady rewards, allowing more adaptive and robust policy to volatile environment (e.g., various contents). More generalization comparisons with other schemes are provided in the Appendix.

\textbf{Complexity Analysis}: 
We compare the computational complexity of our RL-based scheme with existing methods \cite{li2022rate, zhang2023neural, chen2023sparse}, using the real-time NVC method DCVC-RT as the baseline. The evaluation considers multiple metrics, including network parameter count (M), computational cost measured in KMACs per pixel, memory usage (GB), and encoding/decoding throughput (FPS, Frames Per Second). Benefited from the lightweight network design, our proposed method introduces minimal computational overhead—only an additional 0.57 M parameters, 1.60 KMACs per pixel, and 0.33 GB of memory compared to the baseline. Furthermore, by incorporating a down-sampling operation, our method can even improve encoding and decoding throughput. In contrast, other methods tend to compromise real-time performance due to either heavy network architectures \cite{zhang2023neural} or additional pre-coding steps \cite{chen2023sparse}. Although \citet{chen2023sparse} also adopts a down-sampling strategy, their method requires pre-coding equidistant frames with all candidate parameters to initialize the model, which incurs substantial extra encoding time. These results highlight the efficiency and practicality of our approach, demonstrating its potential for real-world deployment without sacrificing performance.

% Metrics include network Parameter count (M), KMACs per pixel, Memory usage (G) and Encoding/ Decoding time (s). Benefited from lightweight network design, our method performs negligible computational overhead increase over baseline model, with only extra 0.57 M parameters, 1.60 KMACs per pixel and 0.33 G memory. Moreover, due to the introduction of down-sampling operation, the encoding and decoding time of our method decreases by 19.7\% and 26.5\%, respectively. In contrast, \citet{zhang2023neural} only adjusts the Lagrange multiplier ($\lambda$), leading to no speed improvement. While \citet{chen2023sparse} also applies down-sampling, their method requires pre-coding equidistant frames with all candidate parameters to initialize their model, costing mass additional encoding time. These results highlight the efficiency of our approach, making it a practical solution for NVC while maintaining superior performance.

\begin{table}[tbp]
\centering
\renewcommand{\arraystretch}{1.0}
\resizebox{0.45\textwidth}{!}{
\begin{tabular}{lcccccc}
\toprule
\multirow{2}{*}{\bf Method} 
& \multicolumn{3}{c}{\bf Complexity} & \multicolumn{2}{c}{\bf Throughput (FPS)} \\ 
& {\bf Params.} & {\bf KMACs/pxl} & {\bf Mem.} & {\bf Enc.} & {\bf Dec.} \\
\midrule
Baseline & 66.33 & 421.31 & 2.27 & 102 & 95 \\
\midrule
 Zhang et al. & +2.12 & +6.40 & +1.22 & 68 & - \\
 Chen et al. & - & - & - & 54 & 108 \\
Ours & +0.57 & +1.60 & +0.33 & \textbf{111} & \textbf{109} \\
\bottomrule
\end{tabular}
}
\begin{tablenotes}
    \footnotesize
    \item ``–'' indicates no change compared to the baseline.
\end{tablenotes}
\vspace{-0.5em}
\caption{Complexity Comparison over DCVC-RT}
\label{tab:efficiency}
\end{table}

\subsection{Ablation Studies} \label{sec:ablation}
To further understand the effectiveness of our scheme, we conduct a series of in-depth evaluations. Unless otherwise specified, DCVC is used as the baseline, and comparisons are made with the learned method \cite{zhang2023neural}.

\textbf{Training Frame Numbers:} 
To evaluate our scheme's ability to capture inter-frame dependencies, we train the model with varying frame numbers. As shown in Table~\ref{tab:different_lengths}, increasing training frame numbers steadily improves both R\text{-}D performance and rate control accuracy. In contrast, \citet{zhang2023neural} reports no clear benefits from longer training sequences in their own ablation study. A potential reason is that the deviation caused by not considering inter-frame rate dependencies gradually as frame numbers increase.

\begin{figure}[htbp]
\centering
\includegraphics[width=1.0\linewidth]{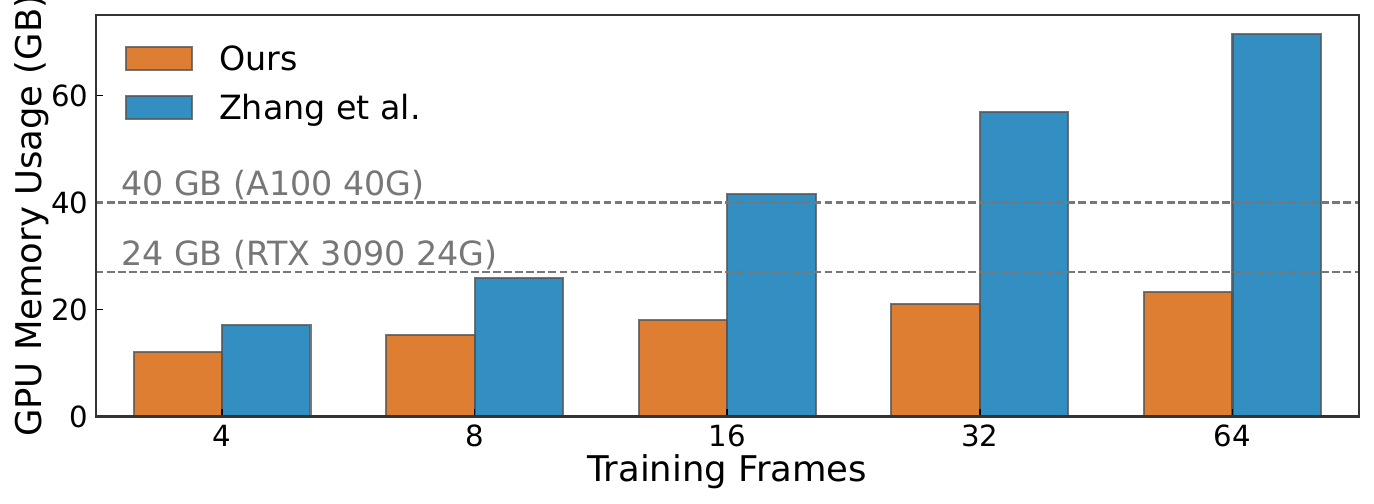}
\vspace{-1.5em}
\caption{Comparison over the GPU Memory Usage with Increasing Training Frames. The memories of RTX 3090 and A100 are marked with dashed lines.}
\label{fig:cuda_memory}
\end{figure}

By contrast, our method models inter-frame dependencies in a frame-wise manner, which accurately models per-frame rate and distortion dependencies, naturally scaling to longer sequences. Furthermore, as shown in Fig.~\ref{fig:cuda_memory}, our approach maintains linear training complexity with respect to sequence length, ensuring both computational efficiency and feasibility.

% To assess our scheme's ability to capture inter-frame dependencies, we train our model with varying frame numbers. As shown in Table \ref{tab:different_lengths}, increasing training frames gradually improves both $R\text{-}D$ performance and rate control accuracy. In contrast, \citet{zhang2023neural} fail to benefit from increased training frames, as reported in their own ablation study. A potential reason is that such window based methods suffer from more difficult rate-distortion-coding parameters as the number of frames grows, making optimization and convergence intractable. 

% In contrast, our approach independently models inter-frame dependencies frame by frame, enabling better scalability. Besides, as shown in Fig.~\ref{fig:cuda_memory}, our method can still maintain linear training complexity with increasing frame numbers, ensuring both efficiency and stability.

\begin{table}[htbp]
\centering
\small
\resizebox{0.4\textwidth}{!}{
\begin{tabular}{@{}lcccccc@{}}
\toprule
\ & 4 & 8 & 16 & 32 & 64 \\ 
\midrule
BD-Rate (\%) & -8.84 & -11.15 & -15.03 & -16.49 & -16.90 \\ 
$\Delta R$ (\%) & 2.48 & 1.82 & 1.67 & 1.48 & 1.43 \\ 
\bottomrule
\end{tabular}
}
\vspace{-0.5em}
\caption{Results with Different Training Frame Numbers}
\label{tab:different_lengths}
\end{table}

\textbf{Performance Under the Setup of \citet{zhang2023neural}:} 
In~\citet{zhang2023neural}, the mini-GOP size is fixed at 4, and only the Lagrange multiplier $\lambda_t$ is used for rate control. To ensure a fair comparison, we replicate this setup. As shown in Table~\ref{tab:retrain}, our method consistently outperforms~\citet{zhang2023neural} across multiple datasets, demonstrating its superiority brought about by considering inter-frame rate and distortion dependencies, even under identical setups.

% In \citet{zhang2023neural}, the window (mini-GOP) size is fixed at 4, and only $\lambda_t$ is used for rate control. To ensure a fair comparison, we align our method with them. As shown in Table \ref{tab:retrain}, our method consistently outperforms \citet{zhang2023neural} across various datasets, demonstrating superior performance even under identical conditions.

%This superiority stems from the cumulative reward and trajectory optimization of RL. 
\begin{table}[ht]
    \centering
    \renewcommand{\arraystretch}{0.9}
    \setlength{\tabcolsep}{1.43pt} 
    \small
    \resizebox{0.44\textwidth}{!}{ 
    \begin{tabular}{lcccccc}
        \toprule
        \textbf{Metric} & \textbf{UVG} & \textbf{MCL-JCV} & \textbf{Cls.B} & \textbf{Cls.C} & \textbf{Cls.D} & \textbf{Cls.E} \\
        \midrule
        BD-Rate (Zhang et al.)     & -7.34 & -5.68 & -5.88 & -4.42 & -3.80 & -9.24 \\
        BD-Rate (Ours)                         & -7.74 & -6.60 & -7.32 & -5.22 & -4.16 & -11.05 \\
        \midrule
        $\Delta R$ (Zhang et al.)  & 2.80 & 2.95 & 2.32 & 1.94 & 2.11 & 1.33 \\
        $\Delta R$ (Ours)                      & 2.58 & 2.44 & 1.98 & 1.91 & 1.80 & 1.07 \\
        \bottomrule
    \end{tabular}
    }
    \vspace{-0.5em}
    \caption{Comparison with \citet{zhang2023neural} with Only $\lambda_t$}
    \label{tab:retrain}
\end{table}

\textbf{Impact of Down-Sampling Factors $m_t$:} 
We further investigate the contribution of jointly deciding both $\lambda_t$ and the down-sampling factor $m_t$. As shown in Table~\ref{tab:impact_M}, while both strategies achieve accurate rate control, using only $\lambda_t$ yields limited PSNR gains, particularly at high bitrates. In contrast, jointly optimizing $\lambda_t$ and $m_t$ enables a richer R\text{–}D trade-off space, leading to significantly improved overall performance.

% We further analyze the $R\text{-}D$ performance by comparing (i) deciding only $\lambda_t$ and (ii) joint deciding $m_t$ and $\lambda_t$. As shown in Table~\ref{tab:impact_M}, while both strategies achieve accurate rate control, relying solely on $\lambda_t$ offers limited PSNR gains, especially at high rate range. In contrast, joint $\lambda-m$ selection provides a richer set of $R\text{-}D$ candidates, leading to lower distortion.

\begin{table}[ht]
\centering
\renewcommand{\arraystretch}{1.1}
\small % 调整字体大小为小号
\resizebox{0.45\textwidth}{!}{ % 缩放表格到0.5页宽
\begin{tabular}{ccccccc}
\toprule
\multirow{2}{*}{$\lambda_t$} & \multicolumn{2}{c}{Baseline} & \multicolumn{2}{c}{w/ $\lambda_t$} & \multicolumn{2}{c}{w/ $m_t$ and $\lambda_t$} \\
\cmidrule(lr){2-3} \cmidrule(lr){4-5} \cmidrule(lr){6-7}
 & BPP & PSNR & BPP & PSNR & BPP & PSNR \\
\midrule
256  & 0.0251 & 33.502  & 0.0249 & 34.138  & 0.0250 & 34.306 \\
512  & 0.0489 & 35.329  & 0.0489 & 35.582  & 0.0488 & 35.827 \\
1024 & 0.0710 & 36.248  & 0.0709 & 36.344  & 0.0711 & 36.560 \\
2048 & 0.1006 & 37.028  & 0.1005 & 37.052  & 0.1001 & 37.245 \\
\bottomrule
\end{tabular}
}
\vspace{-0.5em}
\caption{Ablation Results Regarding The Impact of $m_t$}
\label{tab:impact_M}
\end{table}

\section{Conclusion}\label{sec:conclusion}
% Despite the accepted benefits of more extensive reference, existing model-based RC paradigms are constrained by the assumption of temporal stationarity and exponential complexity, which is negative for NVC. 
In this paper, we revisited the rate control problem in NVC and highlighted the importance of jointly modeling inter‐frame rate and distortion dependencies. To this end, we proposed a reinforced rate‐control framework that accurately models these environmental conditions as states, learns a dynamic policy to directly map them to per‐frame coding parameters, and optimizes via expected cumulative discounted return. Extensive experiments—covering multiple codecs and diverse settings—show that explicitly incorporating inter‐frame rate and distortion dependencies significantly reduces rate error, enhances R\text{-}D performance, improves generalization, and maintains low computational complexity. These advancements position our method as a practical solution for real-world NVC deployments, particularly in bandwidth-constrained or dynamic network environments. Future work will integrate network transmission conditions (e.g., packet loss and congestion) to develop network‑aware rate control schemes for NVC that jointly optimize performance and network adaptability.

\section*{Acknowledgments}
This work was supported in part by Natural Science Foundation of Jiangsu Province (Grant No. BK20241226, BK20243038) and Natural Science Foundation of China (Grant No. 62401251, 62431011, 62471215, 62231002). The authors would like to express their sincere gratitude to the Interdisciplinary Research Center for Future Intelligent Chips (Chip-X) and Yachen Foundation for their invaluable support.

\bibliography{aaai2026}

\clearpage
\appendix
\section{More implementation details} 
\label{appendix:Appendix.A}
\textbf{Metrics}: To assess the quality of reconstructed frames, we use Peak Signal-to-Noise Ratio (PSNR), while the bitrate efficiency is measured using bits per pixel (BPP). The overall rate-distortion ($R\text{-}D$) performance is evaluated using the widely used Bjontegaard-Delta bitrate (BD-Rate)  \cite{bjontegaard2001calculation} metric. Additionally, the accuracy of rate control is measured by the absolute percentage difference between the remaining bitrate $R_{rem}$ and the target bitrate $R_{tar}$, i.e., $\Delta R= \frac{\lvert R_{rem} \rvert}{R_{tar}} \times 100 \%$. Furthermore, since each frame's encoding and decoding requires both $\lambda_t$ and $m_t$, so they should be encoded into the bitstream. $\lambda_t \in \left[256.0, 2048.0\right]$ operates with single decimal place precision, requiring approximately 11 bits (10 bits for the DCVC-DC where the range is [85, 840]). And $m_t \in \left[0.5, 1.0\right]$ with two decimal place precision, demands about 6 bits.

\textbf{Codec baselines}: We believe that our three codec baselines exhibit sufficient differentiation and comprehensive differences in temporal dependencies and rate-distortion characteristics, which may technically come from different paradigms of residual coding or conditional coding, pixel domain reference or feature domain reference, different variable rate implementation methods, hierarchical quality modulation, repetitive compression training, different intra frame coding methods, etc. In addition, for the variable rate fine-tuning of DVC and DCVC, we used the Vime90K dataset and a maximum of 7 frames for training. From Fig. \ref{fig:rd_curve}, it can be seen that the performance of our trained codecs are basically consistent with the original ones.

\textbf{Benchmarks}: To compare fairly with prior work, we reproduce the results of existing methods. For \citet{chen2023sparse}, which does not require training, we directly test their approach under the same conditions described in their paper as much as possible. For \citet{zhang2023neural}, we align their training dataset with our mixed dataset, as their original method only used BVI-DVC. While our reproduction of \citet{chen2023sparse} yields results close to the reported ones, our reproduction of \citet{zhang2023neural} shows slightly lower performance, which we verified was not due to dataset differences. For a fair comparison, we use raw
experimental data from their original papers. Some additional results are based on our faithful implementations due to unavailable sources.

\textbf{Stable training tricks}: For both training codec and our RL-based rate control scheme, the gradients of consecutive frames in temporal sequences exhibit inherent correlations, which may lead to outlier gradients or unstable gradient conflicts. To address this issue, we leverage several stabilization techniques during training, including gradient clipping and regularization, warm-up and cosine-annealed learning rate decay schedule, validation dataset monitor, and parameter soft-updating~\cite{haarnoja2018soft, haarnoja2019softactorcriticalgorithmsapplications}. These mechanisms collectively alleviate the negative impact of premature training strategies on model convergence. 

\begin{figure*}[ht]
\centering
\subfloat[]{\includegraphics[width=0.45\linewidth]{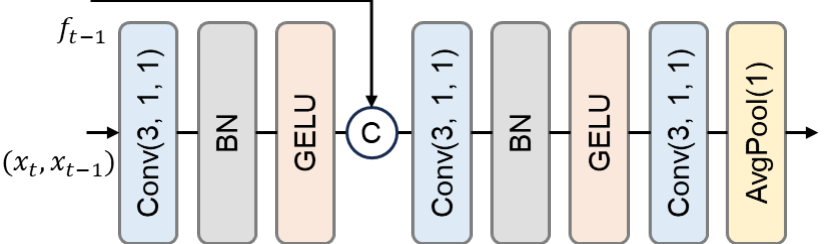}}
\hspace{0.5cm}
\subfloat[]{\includegraphics[width=0.215\linewidth]{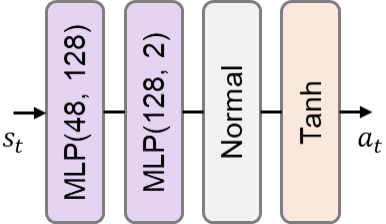}}
\hspace{0.5cm}
\subfloat[]{\includegraphics[width=0.15\linewidth]{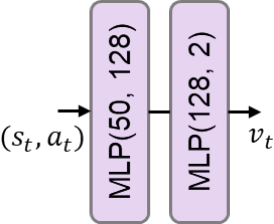}}
\caption{{\bf Designs of networks.} (a) Feature extractor, (b) Actor network, (c) Critic network.}
\label{fig:network_design}
\end{figure*}

\textbf{Algorithm}: The training and inference processes of our reinforcement learning-based rate control method are detailed in Algorithm \ref{alg:training_algorithm} and Algorithm \ref{alg:inference_algorithm}. The architecture of the Actor network, Critic network, and State Feature Extractor is illustrated in Fig. \ref{fig:network_design}, while Table \ref{tab:hyperparameters} presents the hyperparameter configurations. The composition of the state space is further detailed in Table \ref{tab:state_definition}.

\begin{algorithm*}
\caption{Training Process of Our Reinforcement Learning Based Rate Control Algorithm}
\label{alg:training_algorithm}
\begin{algorithmic}[1]
\setlength{\itemsep}{0pt} 
\STATE \textbf{Initialize} Hyperparameters: reward discount factor $\gamma$, weight of distortion term reward $\delta$, weight of bitrate term reward $\eta$, soft update rate $\xi$, entropy weight $\epsilon$; learning rate $\alpha$, $\beta$, $\kappa$ for Actor network, Critic network and feature extractor
\STATE \textbf{Initialize} Actor network's params $\theta$, Twin Critic networks' params $\phi$, feature extractor's params $\psi$
\STATE \textbf{Initialize} Delayed Actor network's params $\tilde{\theta}$, Delayed Twin Critic networks' params $\tilde{\phi}$
\STATE \textbf{Initialize} optimizers for Actor and Critic networks
\STATE \textbf{Initialize} Video dataloader and Replay Buffer (a stack) $\mathrm{RB}$
\FOR{each $epoch$}
    \FOR{each $iteration$}
        \STATE \textbf{Step 1: Obtain trajectory $\tau_i$ by coding}
        \STATE Sample $\mathcal{X}=\{x_0, x_1,..., x_T\}$ from video dataloader
        \STATE Obtain target bitrate and other side information with random $\lambda_t$ and $m_t$ for each frame.
        \FOR{each $x_t \in \mathcal{X}$}   \label{alg:for_t_in_T}
            \STATE Feed $x_t, x_{t-1}, f_{t-1}$ into feature extractor, integrating other side information in Table. \ref{tab:state_definition} to obtain state $s_t$
            \STATE Sample action $a_t$ ($\lambda_t$, $m_t$) with Actor network for each frame $x_t$
            \STATE Encode and Decode $x_t$ with $\lambda_t$ and $m_t$
            \STATE Upsample to original resolution, and calculate $R_t$, $D_t$ and $r_t$
        \ENDFOR
        \STATE Reset Environment
        \STATE Pull $\{(s_t, a_t, r_t, s_{t+1}, end)\}$ into $\mathrm{RB}$ and sample $\tau_i$ from the bottom of $\mathrm{RB}$
        \STATE \textbf{Step 2: Compute target value function}
        \STATE Compute the estimated action value $Q_{\phi}^1$, $Q_{\phi}^2$
        \STATE Sample next action $a_{t+1} \sim \pi_\theta(s_{t+1})$
        \STATE Compute entropy $\log \pi_\theta(a_{t+1}|s_{t+1})$
        \STATE Compute the delayed action value $\tilde{Q}_{\phi}^1$, $\tilde{Q}_{\phi}^2$
        \STATE Compute target Q-value:
        \vspace{-0.5em}
        \[\ \ \ \ \ \ \ \ \ \ \ \ \ \ \ \ \ \ \ \ \ \ \ \ \ \ \ \ \ \ \ \ \ \ \ 
        y_t = r_t + \gamma (1 - d_t) \left(\min(\tilde{Q}_{\phi}^1, \tilde{Q}_{\phi}^2) - \epsilon \cdot \log \pi_\theta(a_{t+1} \mid s_{t+1})\right)
        \]
        \vspace{-1.0em}
        \STATE \textbf{Step 3: Update Critic networks}
        \STATE $L_{Q1} = \text{MSE}(Q_{\phi}^1(s_t, a_t), y_t)$
        \STATE $L_{Q2} = \text{MSE}(Q_{\phi}^2(s_t, a_t), y_t)$
        \STATE Update $\phi$ by minimizing $L_{Q1}$ and $L_{Q2}$
    
        \STATE \textbf{Step 4: Update Actor network} if $iteration \bmod 2 = 0$
        \STATE Generate new action $a_t$ from $\pi_\theta(s_t)$
        \STATE Compute Actor loss: 
        \vspace{-0.5em}
        \[
        L_{\pi} = \mathbb{E}\left[\log \pi_\theta(a_t \mid s_t) - Q_{\theta_1}(s_t, a_t)\right]
        \]
        \vspace{-1.5em}
        \STATE Update $\theta$ by minimizing $L_{\pi}$
        \STATE \textbf{Step 5: Soft update Target networks}
        \STATE $\tilde{\theta}_i \leftarrow \xi \theta_i + (1 - \xi) \tilde{\theta}_i$ for $i = 1, 2$
        \STATE \textbf{Step 6: Update $\psi$ with gradient descent} if $epoch < 50$:
        \vspace{-0.0em}
        \[
        \psi \leftarrow \psi - \kappa \nabla_\psi \mathcal{L}(\psi)
        \]
        \vspace{-2.0em}
    \ENDFOR
\ENDFOR
\end{algorithmic}
\end{algorithm*}

\begin{algorithm}
\caption{Inference Process of Our Reinforcement Learning Based Rate Control Algorithm}
\label{alg:inference_algorithm}
\begin{algorithmic}[1]
\setlength{\itemsep}{0pt} % 减少条目之间的垂直间距
\STATE \textbf{Initialize} Actor network $\pi_\theta$, feature extractor $f_\psi$
\STATE \textbf{Input} Video sequence $\mathcal{X} = \{x_0, x_1, ..., x_T\}$
\FOR{each $x_t \in \mathcal{X}$}
    \STATE Extract state representation $s_t = f_\psi(x_t)$
    \STATE Generate action $a_t = (\lambda_t, m_t)$ from $\pi_\theta(s_t)$ based on greedy principle
    \STATE Encode and Decode $x_t$ with parameters $\lambda_t$ and $m_t$
    \STATE Upsample decoded frame to the original resolution
\ENDFOR
\end{algorithmic}
\end{algorithm}

\begin{table}[ht]
\caption{Hyperparameters setup of our reinforcement learning framework.}
\label{tab:hyperparameters}
\centering
\footnotesize % 调整字体大小为脚注大小，紧凑但清晰
\setlength{\tabcolsep}{3pt} % 减少列间距
\renewcommand{\arraystretch}{1.1} % 减小行间距以紧凑显示
\begin{tabular}{c l c}
\toprule
\textbf{Param.} & \textbf{Description} & \textbf{Value} \\ 
\midrule
$\gamma$  & Reward discount factor                & 0.98 \\ 
$\delta$  & Weight of distortion term reward           & 40 \\ 
$\eta$    & Weight of bitrate term reward              & $20\times \frac{POC}{T}$ (200 for T) \\ 
$\xi$     & Soft update rate              & 0.995 \\ 
$\epsilon$ & Entropy weight               & $\text{5e-4}\xrightarrow[\text{epoch}]{\text{300}}\text{1e-5}$ \\ 
$\alpha$  & Actor network learning rate           & $\text{5e-4}\xrightarrow[\text{epoch}]{\text{300}}\text{5e-5}$ \\  
$\beta$   & Critic network learning rate          & $\text{5e-3}\xrightarrow[\text{epoch}]{\text{300}}\text{5e-4}$ \\ 
$\kappa$  & Feature extractor learning rate       & 1e-4 \\ 
$B$       & Sample batch size                     & 32 \\ 
\bottomrule
\end{tabular}
\end{table}

\begin{table}[ht]
\centering
\caption{State space $S_S$ definition.}
\label{tab:state_definition}
\renewcommand{\arraystretch}{1.0} % 增加行高，使表格看起来更加紧凑
\setlength{\tabcolsep}{6pt} % 增加列间距，让内容更清晰
\small
\begin{tabular}{>{\centering\arraybackslash}m{1.2cm} >{\arraybackslash}m{5.25cm}} % 增大列宽，使内容不挤
    \toprule
    \textbf{Index} & \textbf{Components} \\
    \midrule
    1 & Intra-frame feature \\
    2 & Inter-frame feature \\
    3 & Picture Order Counts (POC) \\
    4 & Remaining bitrate \\
    5 & Target bitrate \\
    6 & Last Lagrangian factor \\
    7 & Last down-sampling factor \\
    8 & Remaining frame numbers in current window \\
    9 & Flag if current frame is the first P frame \\
    \bottomrule
\end{tabular}
\end{table}

\begin{figure}[ht]
\centering
\centering
\includegraphics[width=1.0\linewidth]{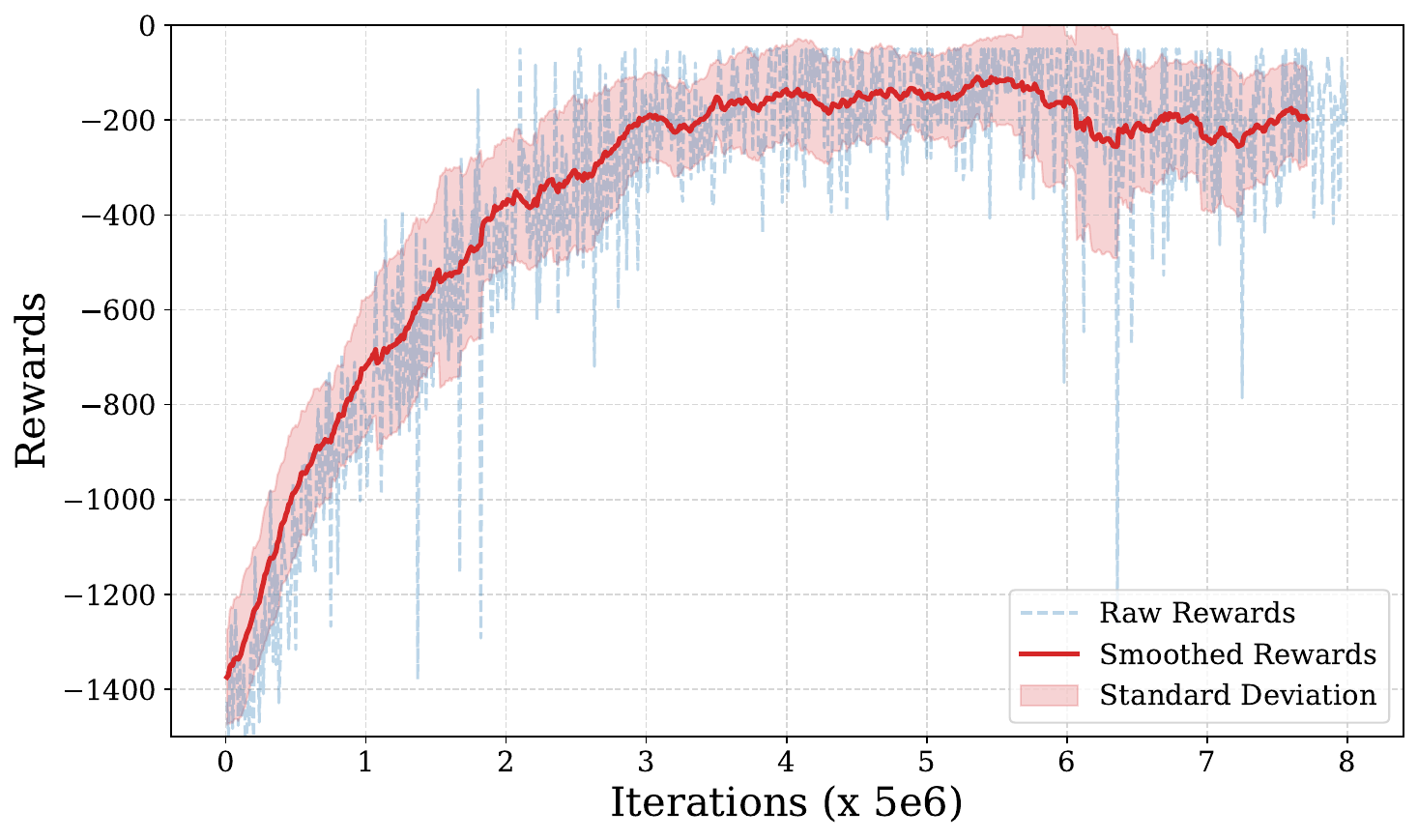}
\vspace{-10pt}
\caption{{\bf The visualization of reward scores curve} during training of our reinforcement learning based rate control method (smoothed). The blue dashed line represents the raw rewards data, the red solid line represents the smoothed one, and the red shadow represents the fluctuation amplitude.}
\label{fig:reward_curve}
\end{figure}

Our method achieves a balanced trade-off between exploration and exploitation, as demonstrated by the reward curve in Fig. \ref{fig:reward_curve}. Initially, high volatility in rewards suggests active exploration of the environment. As training progresses, the model stabilizes and converges towards an optimal rate allocation policy, proving the effectiveness of our approach.

\section{More evaluation results} \label{appendix:Appendix.B}

\begin{figure}[ht]
    \centering
    \subfloat{\includegraphics[width=0.97\linewidth]{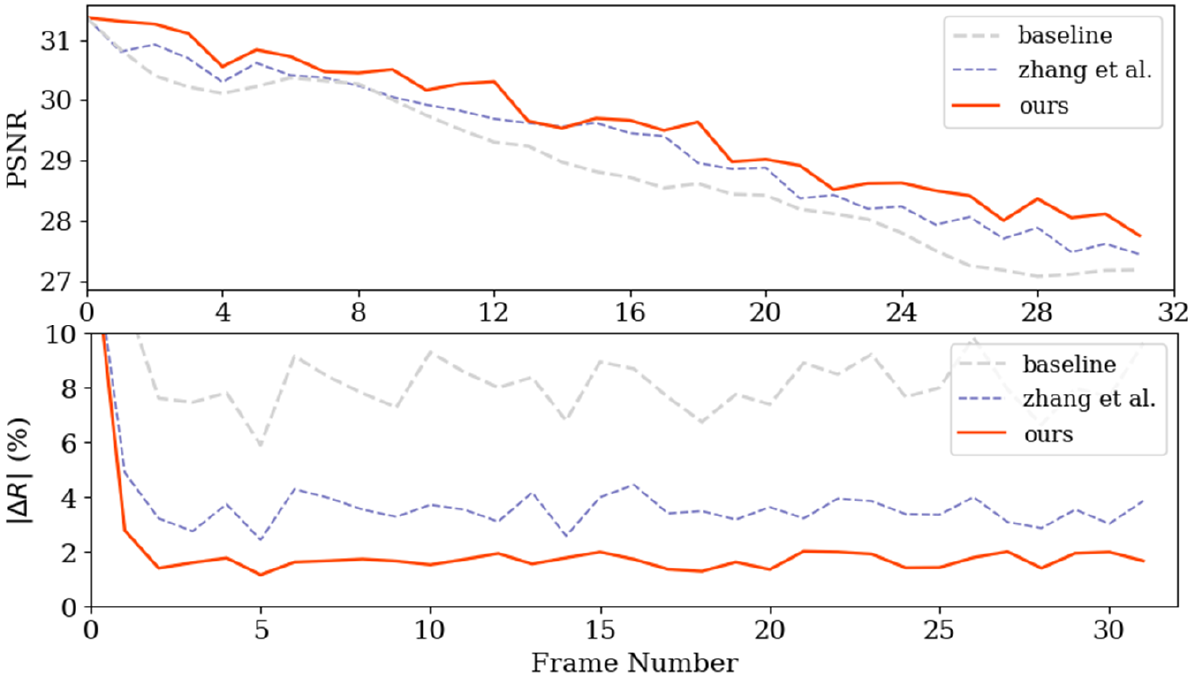}} \\
    % \vspace{-10pt}
    \subfloat{\includegraphics[width=0.99\linewidth]{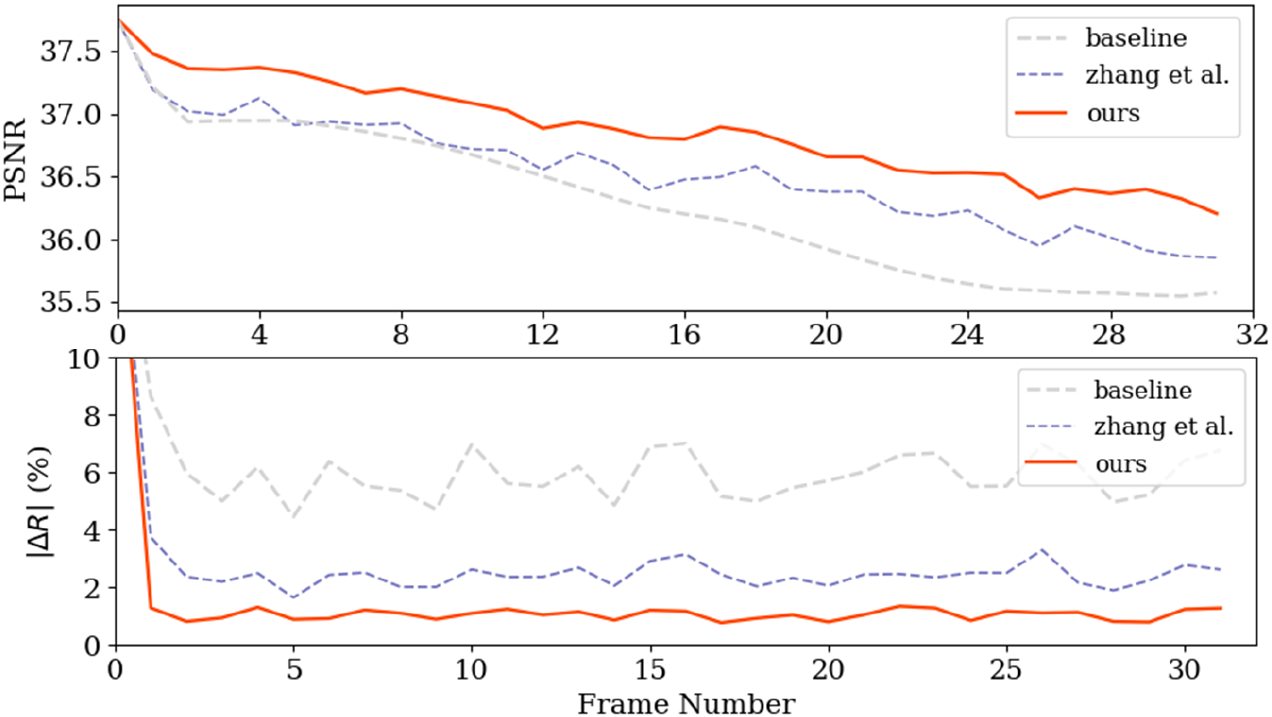}}
    \caption{{\bf More visualizations of rate control effects.} The sequence above is \textit{RaceHorses} at 0.325 BPP, and the sequence below is \textit{Fourpeople} at 0.057 BPP, respectively.}
    \label{fig:more_rd_change}
\end{figure}

\begin{figure}[ht]
\centering
\includegraphics[width=0.98\linewidth]{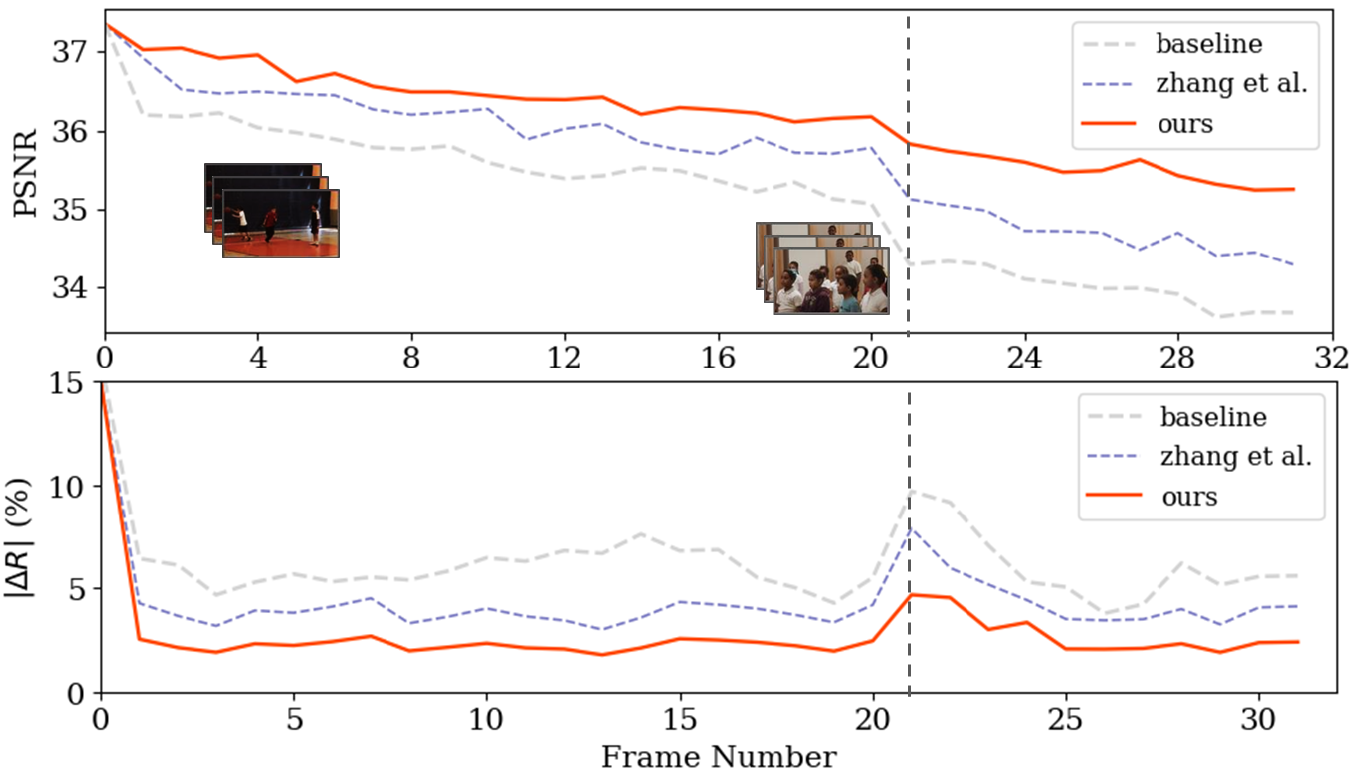}
\caption{{\bf Comparison over a video with scene change}. The content of test sequence will change in the 21st frame, with a sudden drop in quality (above) and an increase in bitrate (below). Our method has a smoother transition when content changes, and gradually eliminates the impact on subsequent frames through dynamic allocation.}
\label{fig:scene_change}
\end{figure}

Figures \ref{fig:rd_curve}, \ref{fig:subjective_quality}, \ref{fig:more_rd_change}, and \ref{fig:scene_change} provide a comprehensive evaluation of our method across various aspects. Specifically, rate-distortion curves are shown on different datasets, subjective quality comparisons illustrate visual improvements, additional visualizations highlight the rate control effect, and evaluations on special sequences further validate robustness. 

\begin{figure}[ht]
\centering
\includegraphics[width=0.98\linewidth]{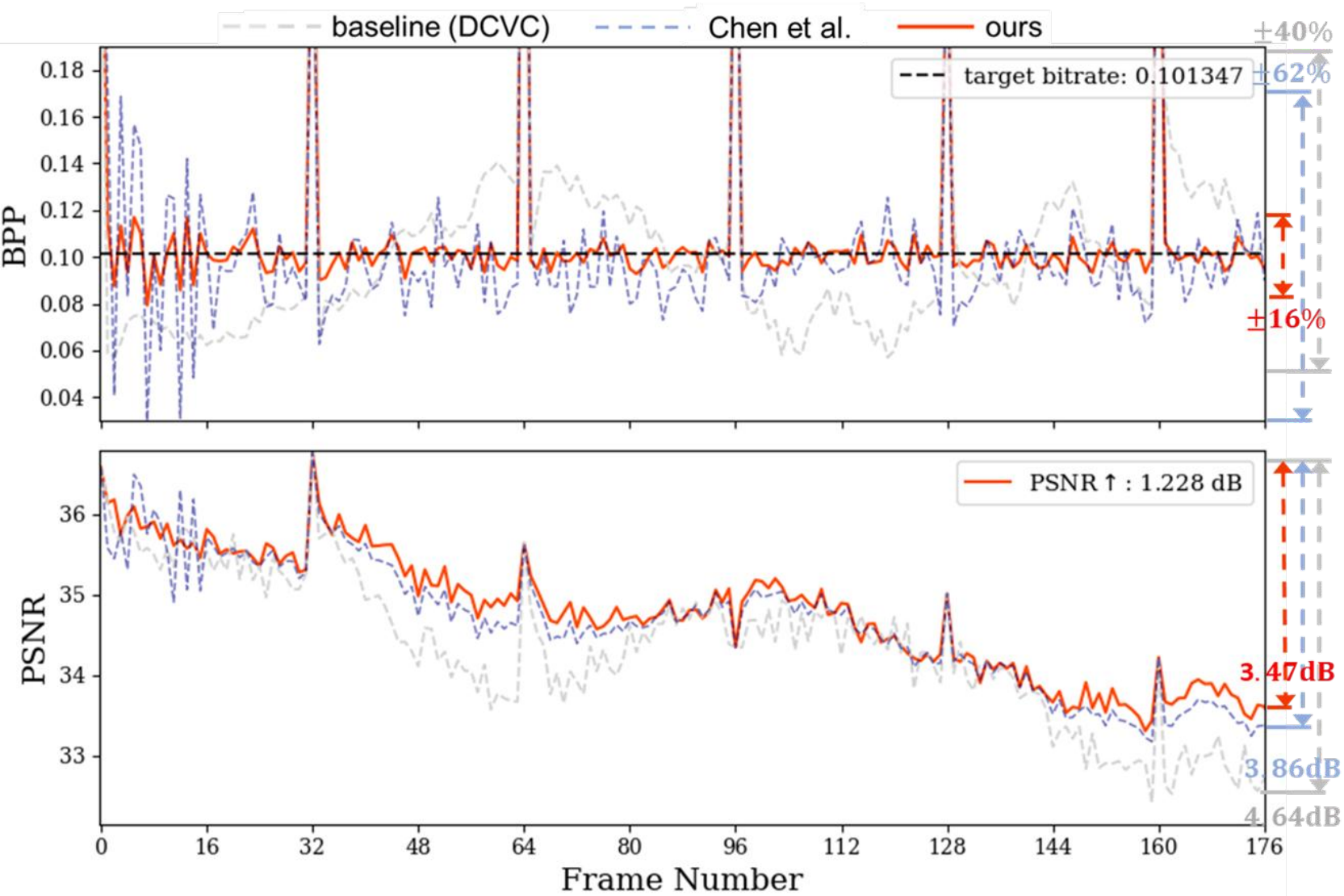}
\caption{{\bf Per-frame rate control visualization with a unified target bitrate \textit{0.101347} BPP on \textit{BasketballDrive} sequence.} (Compared to \citet{chen2023sparse})}
\label{fig:chen_rd_curve}
\end{figure}

To facilitate better comparison, we separately present experimental results against \citet{chen2023sparse}, as shown in Figures \ref{fig:chen_rd_curve}, \ref{fig:chen_rd_curve2}, and \ref{fig:chen_360_video}. We observe that \citet{chen2023sparse} primarily follows a local approximation approach, similar to traditional video codecs. Their method applies uniform bitrate allocation across GOPs and even frames, which is suboptimal for NVC due to its complex temporal reference structures. Additionally, inaccuracies of the $R\text{-}D\text{-}\lambda$ model result in deviations in both bitrate and PSNR, affecting overall rate-distortion performance.

\begin{figure}[t]
\centering
\includegraphics[width=0.95\linewidth]{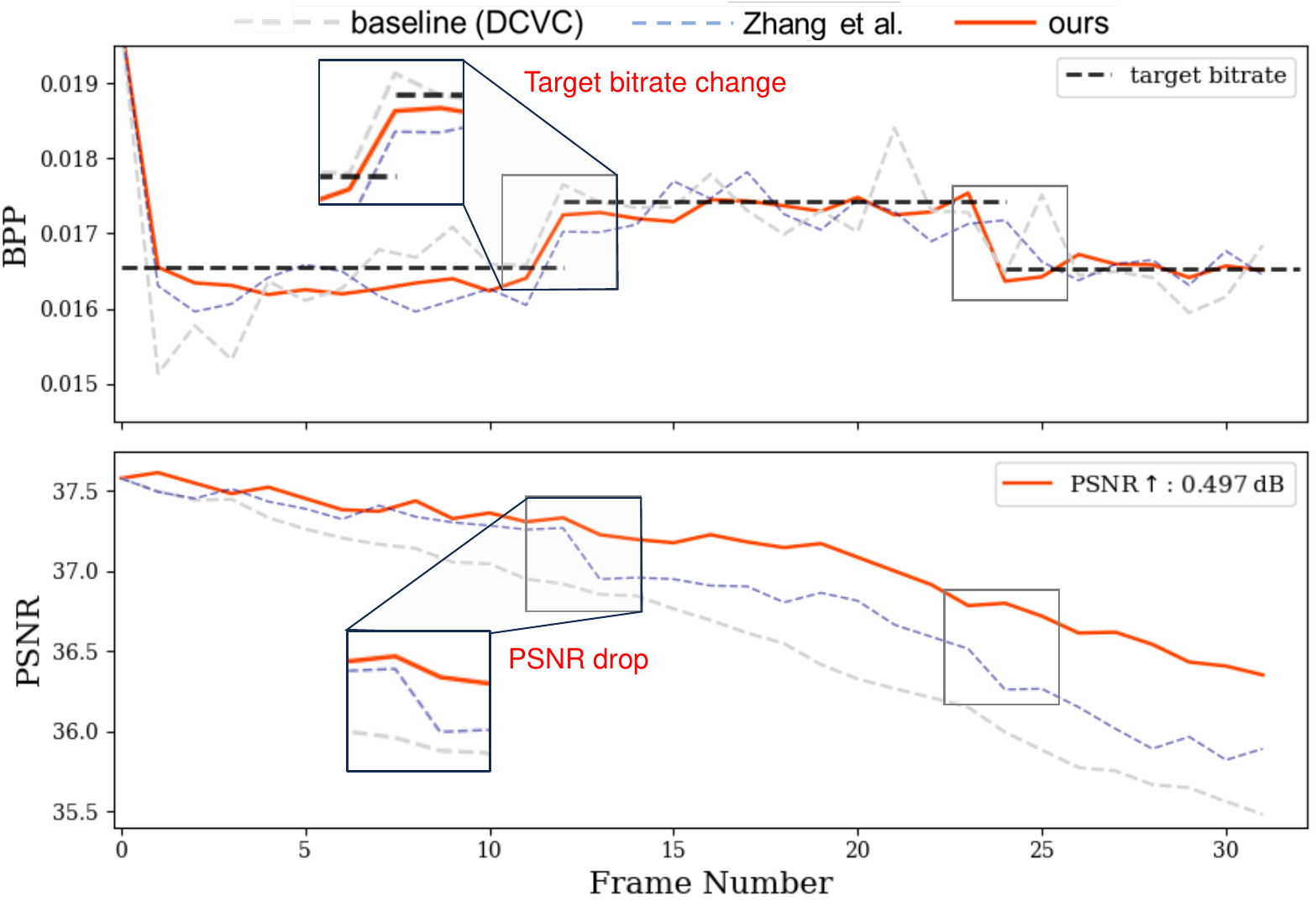}
\caption{{\bf Per-frame rate control effect with a varying target bitrate \textit{0.0163 $\rightarrow$ 0.0173 $\rightarrow$ 0.0166} BPP on \textit{Johnny} sequence.} The rate changes every 0.5 seconds, i.e., 12 frames. (Compared to \citet{zhang2023neural})}
\label{fig:rate_control2}
\end{figure}

\begin{figure}[ht]
\centering
\includegraphics[width=0.98\linewidth]{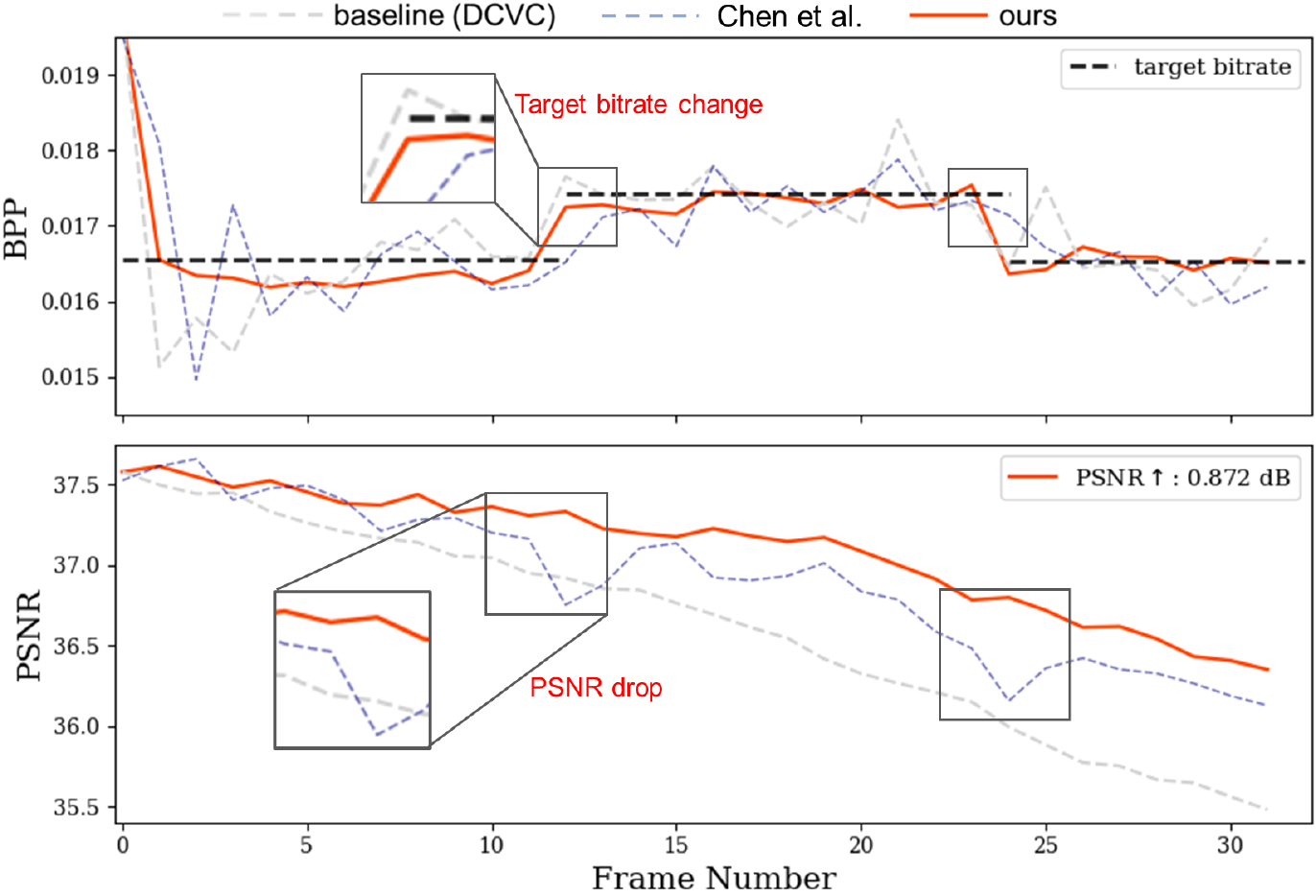}
\caption{{\bf Per-frame rate control visualization with a varying target bitrate \textit{0.0163 $\rightarrow$ 0.0173 $\rightarrow$ 0.0166} BPP on \textit{Johnny} sequence. The rate changes every 0.5 seconds, i.e., 12 frames.} (Compared to \citet{chen2023sparse})}
\label{fig:chen_rd_curve2}
\end{figure}

\begin{figure}[ht]
\centering
\includegraphics[width=0.98\linewidth]{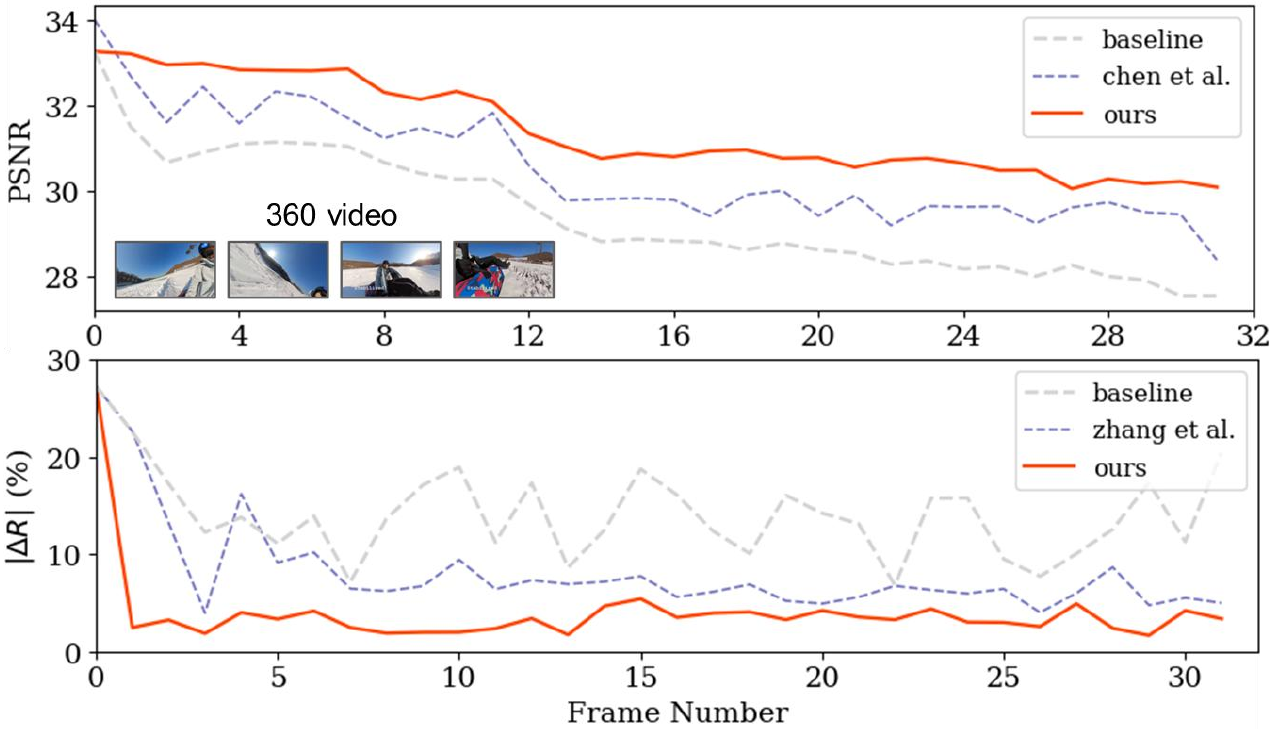}
\caption{{\bf Performance comparison over a 360$^{\circ}$ video}. The test sequence is downloaded from \url{https://www.youtube.com/watch?v=4T8yFnHaJtc}, with the results of quality degradation above and rate fluctuation below. (Compared to \citet{chen2023sparse})}
% ==========================
% if a url is allowed !!!!!!
% ==========================
\label{fig:chen_360_video}
\end{figure}

\section{More ablation studies} \label{appendix:Appendix.C}
\subsection{The effectiveness of learned embedding states} \label{appendix:Appendix.C.1}
To evaluate the effectiveness of our learned embedding states, we replace our feature extractor network with traditional handcrafted metrics used in prior RL-based rate control methods \cite{chen2018reinforcement, zhou2020rate, ho2021dual}. These metrics include the mean and variance of the current frame and residuals, which model intra- and inter-frame characteristics in NVC. However, as shown in Fig.\ref{fig:bad_train}, the training reward curve with traditional metrics exhibits significantly higher variance and lacks a clear convergence trend, even after extensive iterations. This suggests that the temporal reference structure in NVC is a complex combination of prior information across both pixel and feature domains, with cumulative effects over time. Manually designed state struggle to capture this complexity accurately, leading to suboptimal policies and ineffective rate control.

\subsection{The role of each component of reward} \label{appendix:Appendix.C.2}
We further perform an ablation study Eq.~\eqref{eq:reward1} to analyze the effectiveness of our reward design. The expanded form of the inner product consists of three key elements as shown in Eq.~\eqref{eq:reward1} : (i) the $R\text{-}D$ term $r_{rd} = -\delta \cdot D_t$, (ii) the remaining rate penalty term $r_{rem}=- \eta \cdot \frac{\lvert R_{rem} \rvert}{R_{tar}}$ applied per frame with updating $\eta$, and (iii) the RC accuracy term $r_{acc}$ applied to the last frame with fixed large $\eta$. We examine all reasonable combinations of these components, with results summarized in Table \ref{tab:reward_ablation}. We analysis that simultaneously utilizing both $r_{rem}$ and $r_{acc}$ can effectively punish over-allocated situations during training and encourage stably exploring better trade-off between rate accuracy and quality.

\begin{table}[ht]
\centering
\caption{{Ablation studies of reward design.}}
\setlength{\tabcolsep}{3.5pt}
\renewcommand{\arraystretch}{0.75}
\footnotesize
\begin{tabular}{@{}ccc@{}}
\toprule
Reward & BD-rate ($\%$) & \textbf{$\Delta R$ ($\%$)} \\ 
\midrule
$r_{rd}+r_{rem}$       & -4.23       & 3.83       \\
$r_{rd}+r_{acc}$      & -12.97       & 4.79       \\
$r_{rd}+r_{rem}+r_{acc}$ (ours)       & -16.49       & 1.48       \\
\bottomrule
\end{tabular}
\label{tab:reward_ablation}
\end{table}

% \subsection{The role of each component of reward} \label{appendix:Appendix.C.2}
% To analyze the contribution of different reward components, we perform an ablation study by progressively removing or replacing each term, including: (i) the distortion term $r_{d}$, (ii) the remaining rate penalty term $r_{rem}$ applied per frame, and (iii) the rate control accuracy term $r_{acc}$ applied to the last frame (corresponding to large $\eta$ value of the last item). Since the primary objective is to maximize rate-distortion performance while maintaining rate constraints, both the rate-distortion and rate control terms are essential. We examine all possible combinations of these components, with results summarized in Table \ref{tab:reward_ablation}. This analysis highlights the necessity of each term in achieving stable and effective rate control performance.

% \begin{table}[ht]
% \centering
% \caption{{Ablation studies of reward design.}}
% \begin{tabular}{@{}ccc@{}}
% \toprule
% Reward & BD-rate ($\%$) & \textbf{$\Delta R$ ($\%$)} \\ 
% \midrule
% $r_{d}+r_{rem}$       & -4.23       & 3.83       \\
% $r_{d}+r_{acc}$      & -12.97       & 4.79       \\
% $r_{d}+r_{rem}+r_{acc}$ (ours)       & -16.49       & 1.48       \\
% \bottomrule
% \end{tabular}
% \label{tab:reward_ablation}
% \end{table}

\begin{figure*}[ht]
\centering
\subfloat[]{\includegraphics[width=0.29\linewidth]{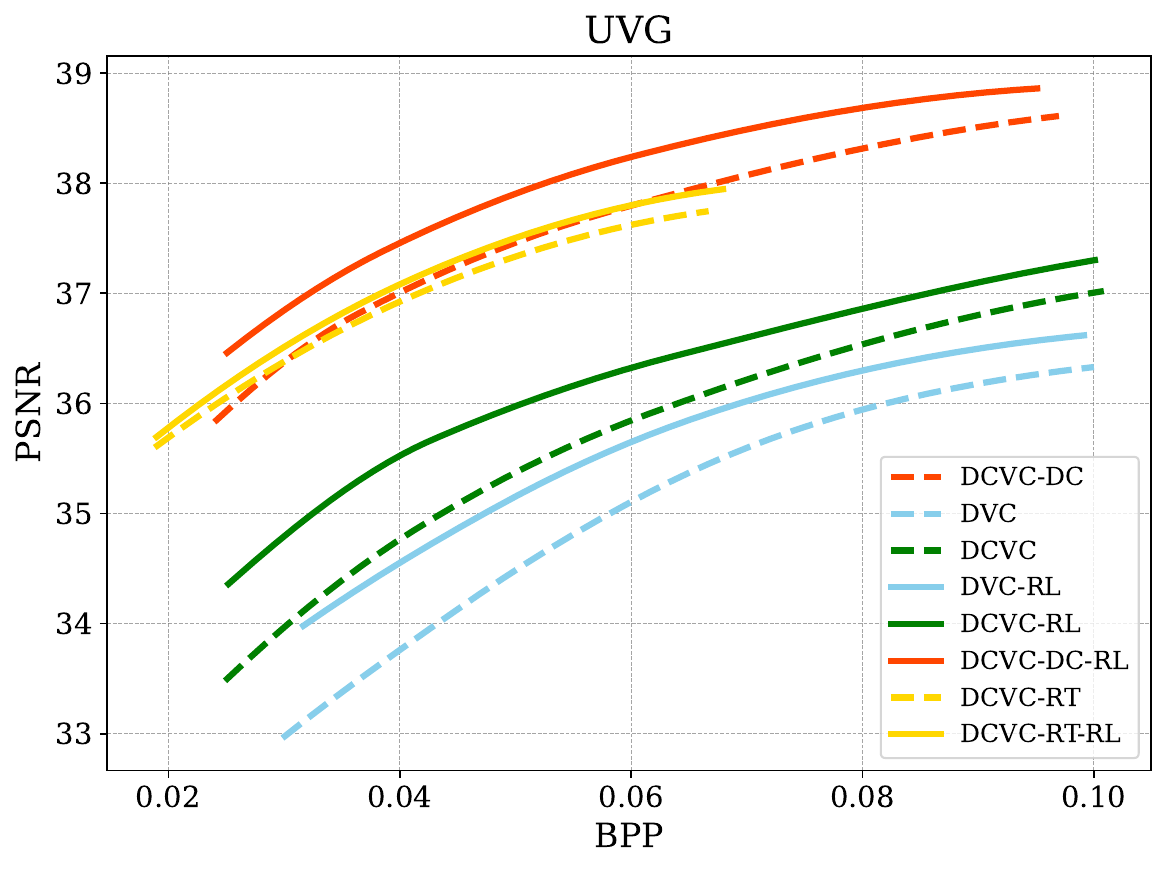}}
\subfloat[]{\includegraphics[width=0.29\linewidth]{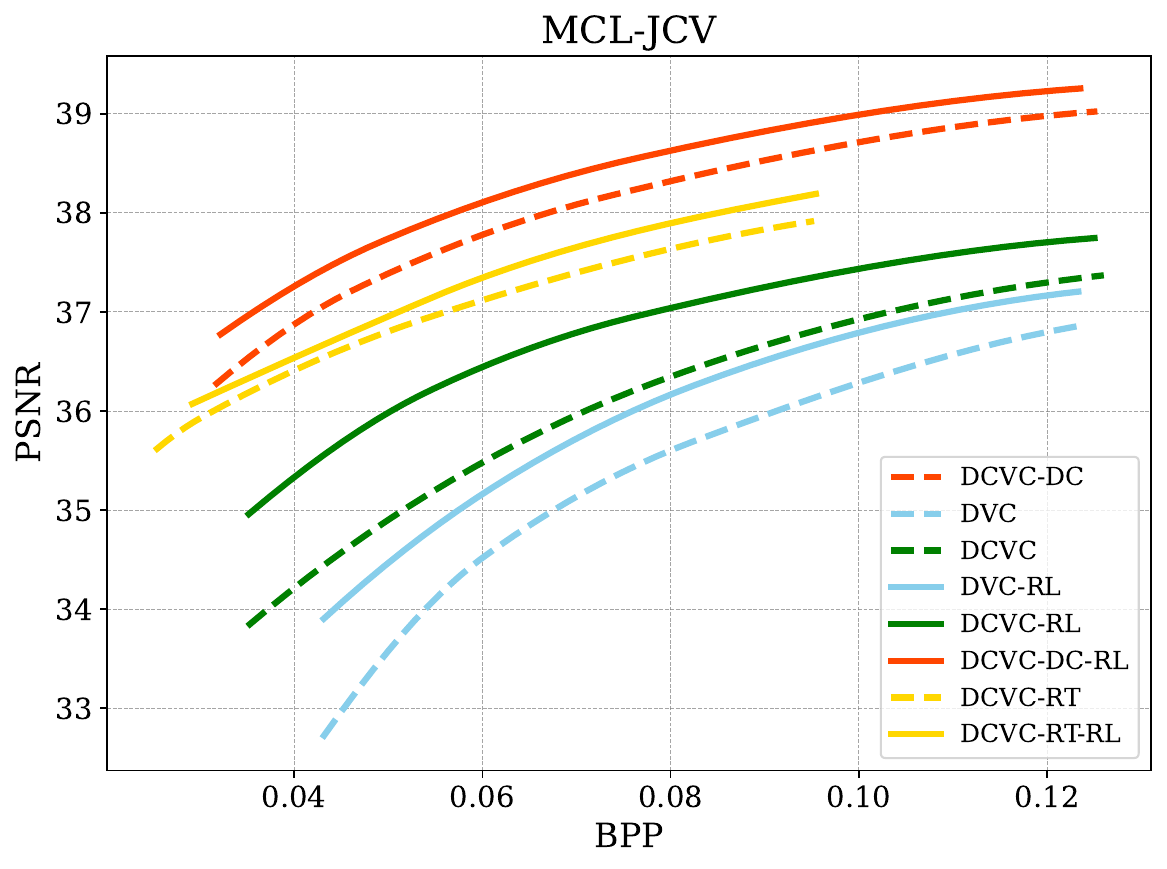}} 
\subfloat[]{\includegraphics[width=0.29\linewidth]{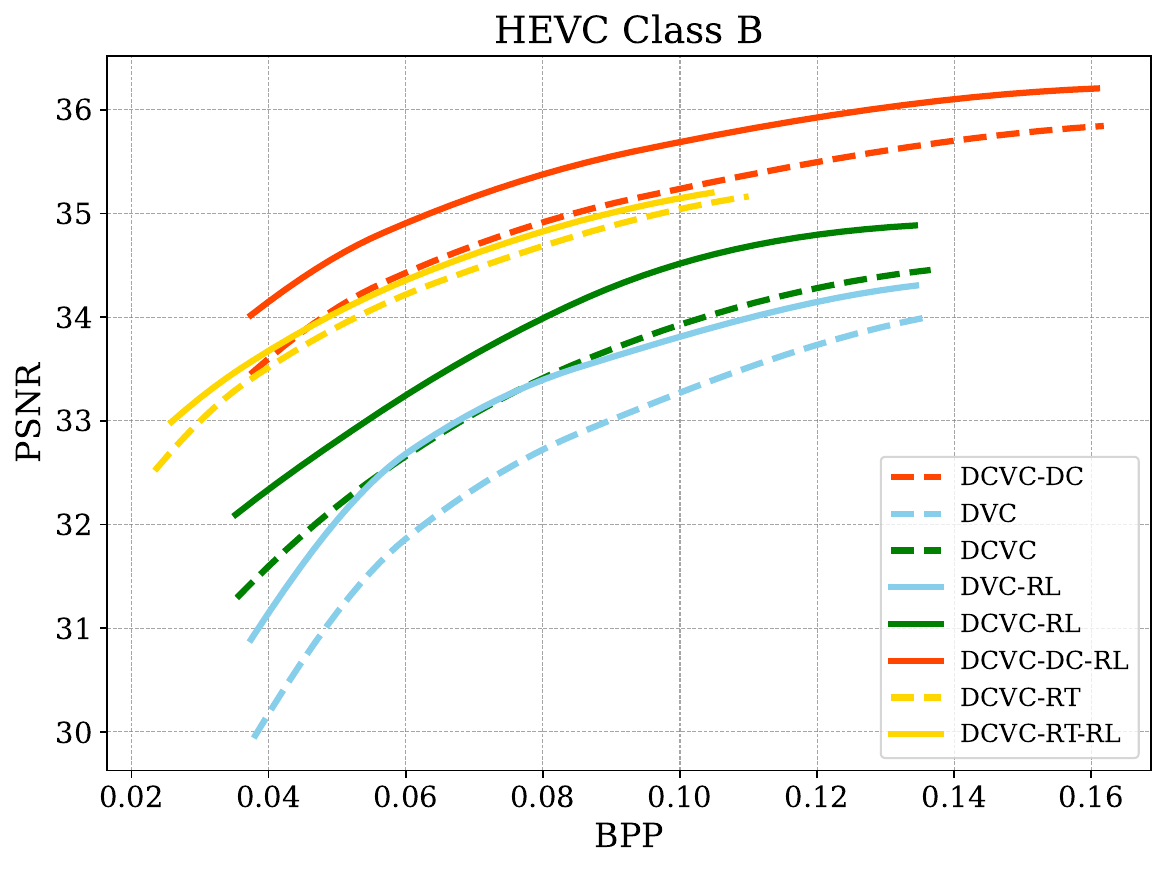}} \\
\vspace{-5pt}
\caption{{\bf Overall $R\text{-}D$ curves.} Solid and dashed lines represent variable-rate models with and without our rate control method.}
\label{fig:rd_curve}
\end{figure*}

\begin{figure*}[ht]
    \centering
    \subfloat{\includegraphics[width=0.825\linewidth]{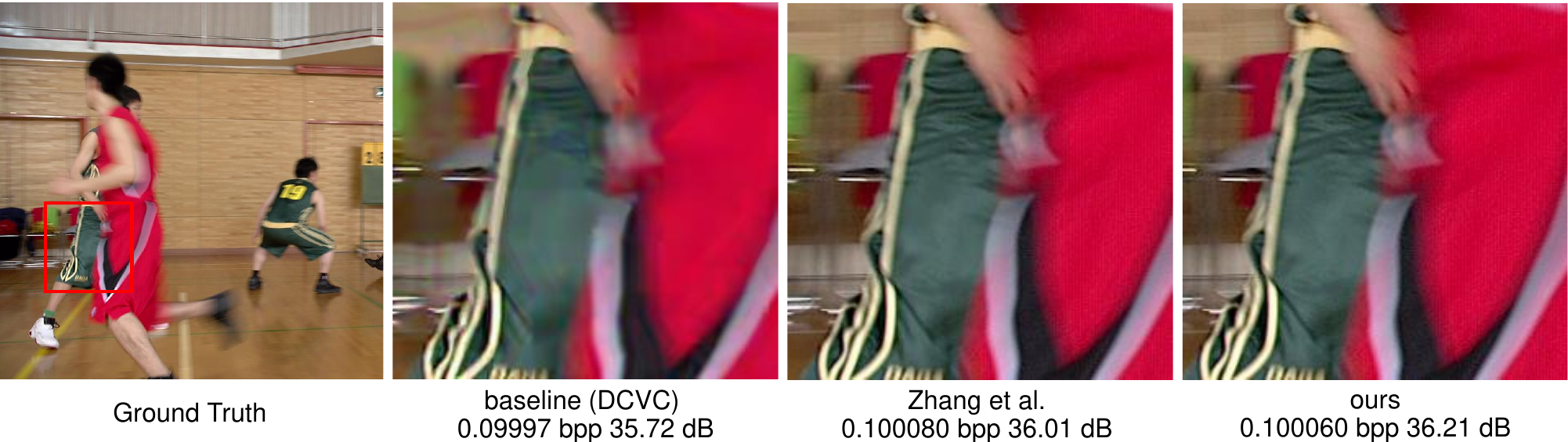}} \\
    % \vspace{-10pt}
    \subfloat{\includegraphics[width=0.825\linewidth]{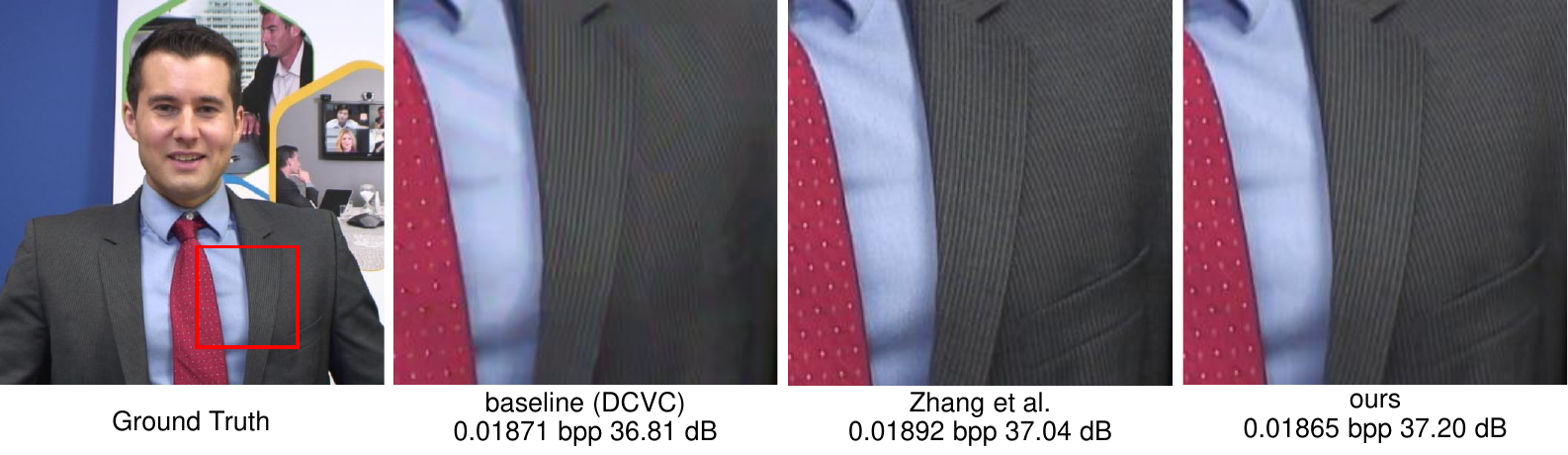}}
    \caption{{\bf Subject quality comparisons.} The sequence above is \textit{BasketballDrive} at 0.1 BPP, and the below is \textit{Johnny} at 0.01875 BPP.}
    \label{fig:subjective_quality}
\end{figure*}

\begin{figure*}[ht]
\centering
\centering
\includegraphics[width=0.5\linewidth]{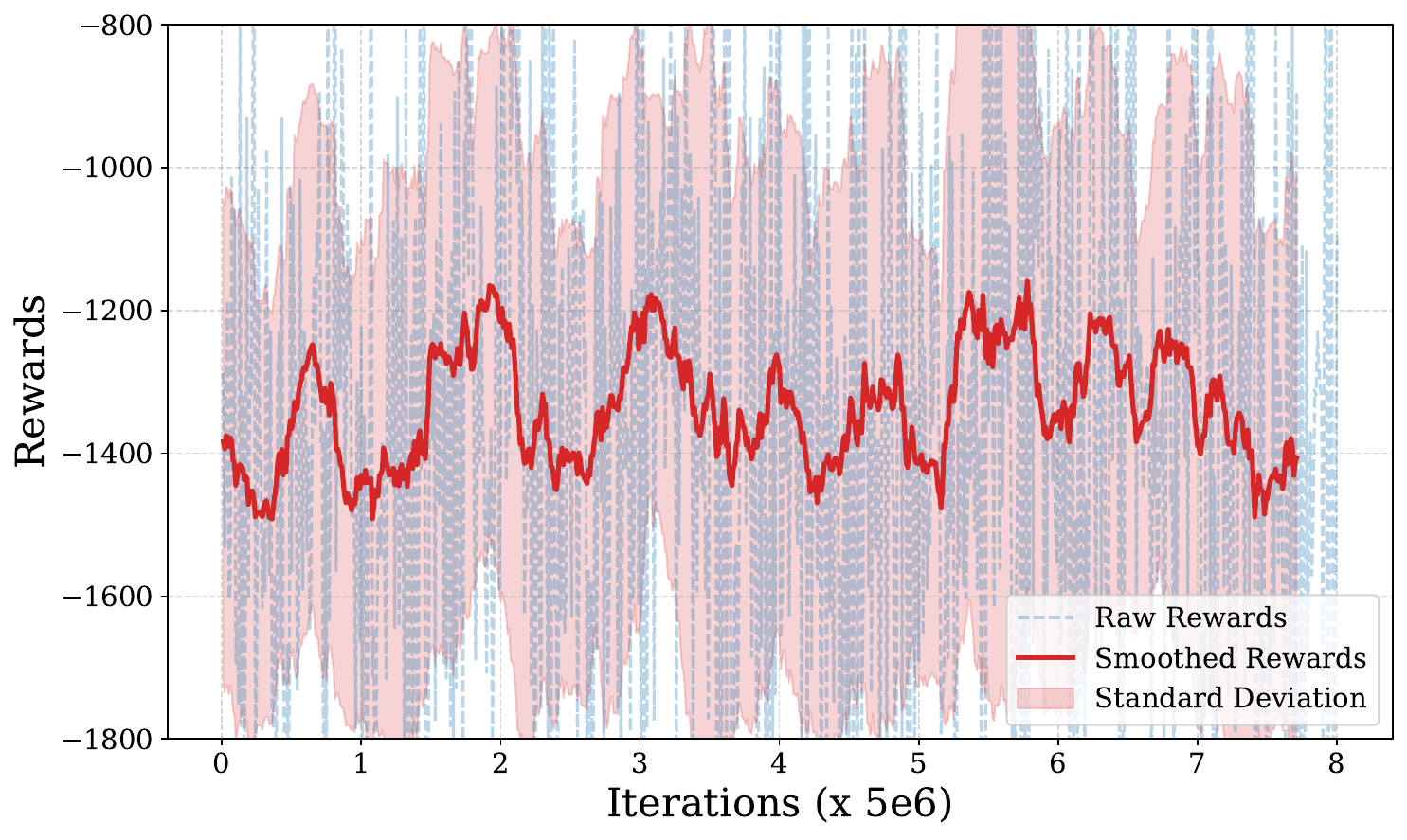}
\caption{{\bf The visualization of reward scores curve with handcrafted state}. The blue dashed line represents the raw rewards data, the red solid line represents the smoothed one, and the red shadow represents the fluctuation amplitude.}
\label{fig:bad_train}
\end{figure*}

\end{document}